\documentclass[pdflatex,sn-nature]{sn-jnl}% Style for submissions to Nature Portfolio journals

\usepackage{graphicx}%
\usepackage{multirow}%
\usepackage{amsmath,amssymb,amsfonts}%
\usepackage{amsthm}%
\usepackage{mathrsfs}%
\usepackage[title]{appendix}%
\usepackage{xcolor}%
\usepackage{textcomp}%
\usepackage{comment}
\usepackage{manyfoot}%
\usepackage{booktabs}%
\usepackage{algorithm}%
\usepackage{algorithmicx}%
\usepackage{algpseudocode}%
\usepackage{listings}%
\usepackage{physics}
\usepackage{siunitx}
\usepackage[english=british]{csquotes}
\newcommand{\fullvec}[1]{\boldsymbol{\vec{#1}}}
\newcommand{\unitvec}[1]{\boldsymbol{\hat{#1}}}

\newcommand{\mat}[1]{#1}
\newcommand{\conf}[1]{\mathrm{#1}}
\newcommand{\operator}[1]{\hat{#1}}
\newcommand{\transpose}[1]{#1^T}
\newcommand{\ctranspose}[1]{#1^H}
\newcommand{\adjoint}[1]{#1^\dagger}

\newcommand{\vary}[1]{\delta_{#1}}
\newcommand{\identity}{\operator{\mathbf{1}}}
\newcommand{\defd}{\overset{\textrm{def}}{=}}
\newcommand{\spinup}{\alpha}
\newcommand{\spindown}{\beta}
\newcommand{\proj}{\operator{P}}
\newcommand{\rej}{\operator{Q}}
\newcommand{\Lagrangian}{\operator{L}}
\NewDocumentCommand{\lagrange}{ o }{\ensuremath{\lambda\IfNoValueF{#1}{_{#1}}}}
\NewDocumentCommand{\expect}{ s m }{%
\IfBooleanTF{#1}{%
\langle #2\rangle%
}{%
\left\langle #2\right\rangle%
}}
\newcommand{\antisym}{\operator{\mathcal{A}}}
\newcommand{\annihilate}[1]{\operator{a}_{#1}}
\newcommand{\create}[1]{\adjoint{\annihilate{#1}}}

\allowdisplaybreaks

\begin{document}

\title{\textbf{Enantiosensitive molecular compass}}


\author*[1]{Philip Caesar M. Flores}\email{flores@mbi-berlin.de}
\author[1]{Stefanos Carlstr\"om}
\author[1]{Serguei Patchkovskii}
\author[1,2,3]{Misha Ivanov}
\author[4]{Vladimiro Mujica}
\author[1,5,6]{Andres F. Ordonez}
\author*[1,3,7]{Olga Smirnova}\email{smirnova@mbi-berlin.de}

\affil[1]{Max-Born-Institut, Max-Born-Str. 2A, 12489 Berlin, Germany}
\affil[2]{Insitute of Physics, Humboldt University zu Berlin, Berlin 12489, Germany}
\affil[3]{Technion - Israel Institute of Technology, Haifa, Israel}
\affil[4]{School of Molecular Sciences, Arizona State University, Tempe, AZ 85287-1604, USA}
\affil[5]{Department of Physics, Imperial College London, SW7 2BW London, United Kingdom}
\affil[6]{Department of Physics, Freie Universit\"at Berlin, 14195 Berlin, Germany}
\affil[7]{Technische Universit\"at Berlin, 10623 Berlin, Germany}

\abstract{
Chirality describes the asymmetry between an object and its mirror image and underlies diverse functionalities across molecular, mesoscopic and bulk matter. A particularly intriguing example is chirality-induced spin selectivity (CISS), where chiral structures generate enantio-sensitive spin polarization. Despite extensive research, its microscopic origin and unexpectedly large magnitude remain unresolved. Here, we isolate the intrinsic coupling between chirality and spin by considering spin-resolved photoionization of randomly oriented chiral molecules under isotropic illumination. We show that electric-dipole photoionization in the presence of spin–orbit coupling generates intrinsic correlations between molecular orientation and photoelectron spin that survive complete isotropic averaging. We identify these spin–orientation correlations as the microscopic origin of CISS in photoionization and reveal their complementary manifestation: selecting the photoelectron spin orients the residual molecular ensemble, realizing spin–orientation locking, whereas selecting molecular orientation produces CISS. Both effects are governed by the same correlation strength, set by the magnitude of a molecular-frame photoionization Bloch vector that defines an enantio-sensitive molecular compass. An analogous compass emerges in photoexcitation. Our results establish spin–orientation correlations as a fundamental ingredient of chiral spin photodynamics and provide a microscopic framework for understanding and exploiting spin selectivity in chiral matter.
}

\maketitle

\section{Introduction}
\label{sec:intro}

Chiral molecular interactions are a remarkable example of a geometrically robust response active in living matter, maintained in complex and noisy environments and operating at ambient conditions \cite{CISS,aiello2022chirality}. It has  potential for applications, e.g., in quantum technologies \cite{gaita2019molecular,atzori2019second,wasielewski2020exploiting,carretta2021perspective,heinrich2021quantum}, with CISS being an  example of useful functionality \cite{chiesa2023chirality}. Originally observed as enantio-sensitive electron spin polarization upon transport through chiral biomolecules \cite{ray1999asymmetric,ray2006chirality,gohler2011spin}, CISS now encompasses a broad range of electronic processes where molecular chirality governs spin polarization \cite{CISS,chiesa2023chirality}. Beyond molecular systems, solid-state materials with chiral crystal lattices exhibit a related but more complex influence on spin: they stabilize noncollinear spin textures such as skyrmion crystals - topologically nontrivial spin configurations whose formation and control are governed by the chirality of the lattice \cite{muhlbauer2009skyrmion,nagaosa2013topological}.

Many CISS-related phenomena are induced by light, suggesting the possibility of ultrafast control over spin-sensitive chemical processes via photoexcitation or photoionization \cite{bloom2024chiral,varillas2026roadmap}. The interplay between spin and chirality naturally suggests spin-orbit coupling (SOC) as the underlying mechanism for spin selectivity. However, several experiments have shown that SOC alone cannot explain the observed effect \cite{kettner2018chirality,mishra2013spin,kettner2016silicon}. Numerous theoretical approaches \cite{nuomin2026theories} including scattering theory \cite{yeganeh2009chiral,medina2012chiral,eremko2013spin,varela2013inelastic}, tight-binding models \cite{gutierrez2012spin,guo2012spin,matityahu2016spin,varela2016effective,peter_yelin_2024}, density functional theory \cite{maslyuk2018enhanced,diaz2018thermal,zollner2020insight}, scattering off magnetic impurities \cite{ghazaryan2020filtering}, electron correlation models \cite{fransson2019chirality}, electron-phonon coupling \cite{fransson2020vibrational,vittmann2023spin}, non-adiabatic coupling \cite{bian2021modeling,teh2022spin}, and field-theoretical treatments \cite{shitade2020geometric} have been proposed, yet a complete understanding remains elusive \cite{CISS}. Diverse challenges for interpretation emerge due to the presence of multiple factors that potentially affect this process in experimental measurements 
thereby complicating its analysis \cite{CISS}, e.g., leads or substrates for molecules, and impurities, defects, or spurious fields for solids.

Here, we examine the simplest yet ubiquitous example of spin-chirality coupling: spin-resolved photoionization of randomly oriented chiral molecules under isotropic illumination, thereby removing any orientational bias introduced by either the well-defined light polarization or initial molecular alignment. 
Once ambiguities related to anisotropic targets or substrates, and complex detection schemes involving leads are removed, the fundamental origins of spin-chirality coupling associated with photodynamics emerge with a striking clarity in an approach which is equally applicable to photoionization and photoexcitation of randomly oriented molecules. 

We find that molecular chirality induces an intrinsic correlation between molecular orientation and photoelectron spin that survives even in a fully isotropic setting. Remarkably, this correlation singles out a unique direction in the molecular frame along which spin and orientation are maximally correlated. 
In phenomenological descriptions of CISS, coupling of the
electron spin to a molecular ``chiral axis'' has been invoked to describe
a preferred spin direction in the molecular frame
\cite{naaman2020chiral}. Here, rather than assuming such an axis, we derive
it directly from the photoionization dynamics: a molecular-frame Bloch
pseudovector %$\fullvec{S}^{M}$ 
defines the direction of maximal
spin--orientation correlation, while its magnitude determines the
correlation strength. It therefore provides an intrinsic,
microscopically defined chiral molecular compass. 
The correlation gives rise to two complementary effects: molecular orientation selects photoelectron spin, revealing CISS, while photoelectron spin selects molecular orientation, leading to the previously unreported phenomenon of enantio-sensitive spin–orientation locking.

The enantio-sensitive molecular compass operates in fields of arbitrary polarization, including fully isotropic polarization. We derive these results below in a fully general and compact approach, and perform \textit{ab initio} calculations for a synthetic chiral system with extremely small ratio ($\simeq0.01$) of spin-orbit interaction to the ionization potential. Remarkably, pronounced locking persists even for spin-unpolarized initial states, with the spin-conditioned mean molecular orientation reaching 0.22 corresponding to an approximate 66:34 head--tail imbalance. The mechanism bypasses the weak magnetic-field interaction, and solely relies on electric-dipole interactions between the light field and the molecule. In closed-shell molecules, photoionization directly links the spin of the cation (hole) to its orientation, establishing an initial spin–orientation correlation that can bias spin-selective phenomena in orientation-sensitive environments (e.g. interfaces or orientation-dependent molecular reactions). Although the hole spin subsequently evolves through spin-orbit and vibronic couplings, this evolution proceeds with an opposite phase in opposite enantiomers. Thus, enantio-sensitive spin-orientation locking triggers enantio-sensitive spin densities and currents in molecular cations. We show that correlations between photoelectron spin and molecular orientations underlying this phenomenon are enabled by chirality.

\section{Results}
\label{sec:results}

\subsection{Photoionization Bloch vector}
\label{subsec:geometric_nature}

We introduce the photoionization Bloch pseudovector and establish its role as the molecular compass governing spin--orientation locking in randomly oriented chiral molecules under isotropic illumination.
For a given molecular orientation $\rho$, the amplitudes of one-photon ionization producing photoelectrons with energy $E=k^2/2$ and spin projections $\mu=\pm 1/2$ are defined by the famous Fermi's Golden Rule:
\begin{equation}
c_{E,\mu}(\rho)
=
\int d\Theta_k^M\,
\fullvec{D}^{M}_{I,\fullvec{k}^{M},\mu}
\cdot
\fullvec{E}^{M}_{p}(\rho).
\end{equation}
Here, $\fullvec{D}^{M}_{I,\fullvec{k}^{M},\mu}$ denotes the molecular-frame
photoionization dipole, $I$ labels the final ionic state,
$\mu=\pm 1/2$ denotes the photoelectron spin projection onto the
molecular-frame quantization axis $\unitvec{\zeta}^{M}$, $\fullvec{E}^{M}_{p}(\rho)$ is the polarization vector of
the ionizing field expressed in the molecular frame for a molecular
orientation $\rho$, and
$\fullvec{k}^{M}$ is the photoelectron momentum in the molecular frame\footnote{We use the superscripts `L' and `M' to denote quantities that are defined in the laboratory and molecular frame, respectively. Quantitites from the molecular frame are related to the laboratory frame via the Euler rotation matrix $R_\rho$, i.e., $\fullvec{a}^L=R_\rho\fullvec{a}^M$.}. 
Since only the photoelectron energy, but not its momentum, is resolved in our setup, the latter is traced out by integrating over all momentum directions, $\int d\Theta_k^M$. 

Formally, the populations and mutual coherence of
the two spin components are encoded in the $2\times2$ density matrix,
whose elements are proportional to
$c_{E,\mu}^*(\rho)c_{E,\nu}(\rho)$.
Tracing over the
molecular orientations $\rho$ and  summing over the unresolved final ionic states we obtain a reduced
spin density matrix for a photoelectron of energy $E$ (see
Methods~\ref{subsec:bloch}):
\begin{equation}
\tilde{\varrho}^{M}_{\mu,\nu}
=
\sum_{I}\int d\Theta_{k}^{M}
\left(
\fullvec{D}_{I,\fullvec{k}^{M},\mu}^{M*}
\cdot
\fullvec{D}_{I,\fullvec{k}^{M},\nu}^{M}
\right).
\label{eq:density_matrix_def}
\end{equation}
The geometric nature of Eq. \eqref{eq:density_matrix_def} is revealed by expressing it as:
\begin{equation}
\frac{\tilde\varrho^{M}}{\mathrm{Tr}[\tilde\varrho^{M}]}
= \frac{1}{2}\left(\,\mathbb{I} + \fullvec{S}^{M}\cdot\unitvec{\sigma}^M\right), 
\label{eq:density_matrix_decomposition}
\end{equation}
where,  $\hat{\boldsymbol{\sigma}}$ is the vector of Pauli spin matrices, $\mathrm{Tr}[\tilde\varrho^{M}]\equiv S_0$, $S_0$ is proportional to the total ionization rate, $\tilde\varrho^{M}/\mathrm{Tr}[\tilde\varrho^{M}]$ is the normalized reduced density matrix, and $\fullvec{S}^{M}$ is the Bloch vector:
\begin{equation}
\fullvec{S}^{M} = \frac{1}{S_0}\mathrm{Tr}\left(\tilde\varrho^{M}\unitvec{\sigma}^M\right) \quad,\quad S_0 =\sum_{I} \left[  \int d\Theta_k^M\left(  \left| \fullvec{D}_{I,\fullvec{k}^M,\frac{1}{2}}^M \right|^2 +  \left| \fullvec{D}_{I,\fullvec{k}^M,-\frac{1}{2}}^M \right|^2 \right) \right]
\end{equation}
Both $\tilde\varrho^{M}$ and $\fullvec{S}^{M}$ only depend on the photoionization (or photoexcitation) dipoles and encode the properties of molecular states. Below, we apply our geometric method \cite{floresPRA2026} to show that the Bloch vector emerges as a key object characterizing enantio-senstive spin-orientation locking.  

\subsection{Spin-orientation locking in randomly oriented chiral molecules} 
\label{subsec:locking}

We now show that one-photon ionization of randomly oriented enantiopure molecules can establish a correlation between molecular orientation and the detected photoelectron spin. %To identify this spin--orientation locking, we seek a molecular vector $\unitvec{e}^L$ whose expectation value remains non-zero within the subensemble of molecular cations conditioned on a given photoelectron spin projection along the laboratory-frame axis $\unitvec{s}^{L}$.
To eliminate any preferred direction imposed by the light, we consider isotropic illumination, averaging over all directions of the linear polarization vector in the laboratory frame:
\begin{align}
\fullvec{E}^L_p = E_\omega^L \left( \sin\theta_p\cos\varphi_p \unitvec{x}^L + \sin\theta_p\sin\varphi_p \unitvec{y}^L + \cos\theta_p \unitvec{z}^L \right),
\label{eq:E_iso}
\end{align}
where $(\theta_p,\varphi_p)$ describe all possible polarization directions.

To determine whether spin detection can induce a net orientation within an initially randomly oriented molecular ensemble, we consider an arbitrary molecular unit vector $\unitvec{e}^{M}$ and evaluate its laboratory-frame expectation value
$\unitvec{e}^{L}(\rho)$ over the ionization probability conditioned on detecting the photoelectron along the laboratory-frame axis
$\unitvec{s}^{L}$, see Methods \ref{subsec:quantify}:
\begin{equation}
 \langle \unitvec{e}^L \rangle_{\unitvec{s}^L}^{\text{iso}} \equiv \dfrac{1}{N} \int d\Theta_p \int d\rho \int d\Theta_k^L \,  \unitvec{e}^L W^{L}(\hat{\mathbf k}^{L},\hat{\mathbf s}^{L},\rho)  = \dfrac{1}{3}(\fullvec{S}^M \cdot \unitvec{e}^M)\unitvec{s}^L%  \dfrac{\nu}{3}|\fullvec{S}^M|\unitvec{s}^L,
\label{eq:spin_orient_lock}
\end{equation}
\begin{equation}
W^{L}(\hat{\mathbf k}^{L},\hat{\mathbf s}^{L},\rho) = \dfrac{1}{2}\sum_{I,\mu,\nu} \left( \fullvec{D}_{I,\fullvec{k}^{M},\mu}^{L*} \cdot \fullvec{E}_p^{L*}\right)\left( \fullvec{D}_{I,\fullvec{k}^{M},\nu}^{L} \cdot \fullvec{E}_p^{L}\right) \left( \delta_{\mu,\nu} + \unitvec{s}^L \cdot \unitvec{\sigma}_{\nu,\mu}^L \right),
\label{eq:ion_rate}
\end{equation}
where, $N = \int d\rho \int d\Theta_p \int d\Theta_s^L \int d\Theta_k^L \, W^{L}(\hat{\mathbf k}^{L},\hat{\mathbf s}^{L},\rho)=S_0|E_\omega^L|^2/6$ is the integrated ionization rate for either spin projection along an arbitrary detection axis, which, owing to isotropy and vanishing net spin polarization, equals one half of the spin-summed ionization rate. Here, $W^{L}(\hat{\mathbf{k}}^{L},\hat{\mathbf{s}}^{L},\rho)$ is the momentum- and spin-resolved ionization rate for a molecule with orientation $\rho$, and the detected spin projection along $\unitvec{s}^{L}$ is selected
by the projector $\frac{1}{2}(\mathbb{I}\pm\unitvec{s}^{L}\cdot\unitvec{\sigma}^{L})$.
The integrations over $\rho$ and $\Theta_p$ average over all molecular orientations and linear-polarization directions, respectively, and the photoelectron momentum direction is unresolved thus its direction is integrated over $\Theta_k^{L}$. The normalization factor $N$ additionally includes integration over all spin-detection axes $\Theta_s^L$
, i.e. over all directions along which the specific photoelectron spin projection may be detected. 

\begin{figure}[t!]
\centering
\includegraphics[width=\textwidth]{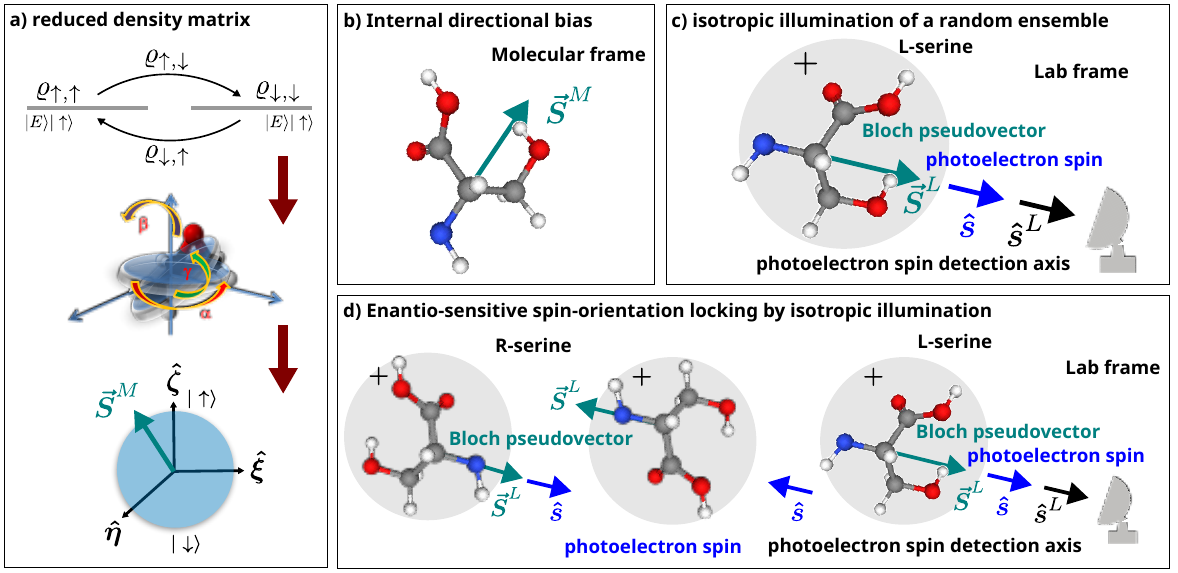}
\caption{Spin-chirality coupling in photoionization of randomly oriented molecules under isotropic illumination is unique to chiral media. (a) The density matrix $\rho$ of a degenerate two-level system corresponding to spin-up $|\uparrow\rangle$ and spin-down $|\downarrow\rangle$ states of the photoelectron with energy $E=\frac{k^2}{2}$ can be averaged over molecular orientations to yield the reduced density matrix $\tilde{\varrho}$. The Bloch vector $\fullvec{S}^M$ is defined on the Bloch sphere in the molecular frame $\{\unitvec{\xi}^M, \unitvec{\eta}^M, \unitvec{\zeta}^M\}$, and is proportional to the expectation value of the spin operator $\unitvec{\sigma}$ for a state with density matrix $\tilde{\varrho}$. (b) The Bloch vector $\fullvec{S}^M$ is ``attached'' to the molecule and represents the molecular compass direction in chiral molecules. (c) After photoionization the cation cloud (gray circle) correlated to photoelectrons with specific spin projection on the spin detection axis $\unitvec{s}^L$ possesses a net orientation such that $\fullvec{S}^L$ is parallel to $\unitvec{s}^L$. 
(d) The direction of spin to cation orientation locking is enantio-sensitive: the photoelectron spin is parallel (antiparallel) to $\fullvec{S}^L$ for right (left) molecules. 
}  
\label{fig:idea_of_locking}
\end{figure}

Equation~(\ref{eq:spin_orient_lock}) shows that, despite averaging over all
molecular orientations and polarization directions, conditioning on the
detected photoelectron spin produces a non-zero net molecular orientation.
For an arbitrary molecular unit vector $\unitvec{e}^{M}$, the magnitude of
this orientation is proportional to its projection onto the photoionization
Bloch vector, $\fullvec{S}^{M}\cdot\unitvec{e}^{M}$. It is therefore maximal
for $\unitvec{e}^{M}\parallel\fullvec{S}^{M}$, identifying the direction of
the Bloch vector as the molecular axis most strongly correlated with the
detected photoelectron spin.

The pseudovector $\fullvec{S}^M$  can be expressed in an equivalent form $\fullvec{S}^M\equiv\nu|\fullvec{S}^M|\unitvec{e}_{\parallel}^{M}$, where $\unitvec{e}_{\parallel}^{M}$ is a unit polar vector along $\fullvec{S}^M$, and $\nu=\pm1$ is a pseudoscalar that distinguishes opposite enantiomers. Defining the angle $\theta$ between an arbitrary molecular vector $\unitvec{e}^{M}$ and  $\unitvec{e}_{\parallel}^{M}$, we can write $\unitvec{e}_{\parallel}^M\cdot\unitvec{e}^M=\cos\theta$ such that 
%We thus define a unit polar vector $\unitvec{e}^{M}$ along this axis. 
the spin-conditioned orientation becomes
\begin{equation}
 \langle \unitvec{e}^L \rangle_{\unitvec{s}^L}
^{\mathrm{iso}}
=
\dfrac{\nu}{3}
|\fullvec{S}^{M}|\cos\theta\,\unitvec{s}^{L}.
\label{eq:microscopic}
\end{equation}
Thus, the photoionization
Bloch pseudovector defines the molecular compass: its direction identifies the
molecular axis $\unitvec{e}_{\parallel}^{M}$ that is maximally locked to the detected photoelectron spin ($\cos\theta=1$),
while its magnitude determines the strength of this spin--orientation
locking. The molecular compass defines the direction of maximal spin-molecular orientation correlations: cations correlated with photo-electrons having positive (negative) spin projection onto the detection axis $\unitvec{s}^L$ are oriented with $\unitvec{S}^L$ parallel (antiparallel) to $\unitvec{s}^L$. This phenomenon, unique to chiral media, ensures opposite orientations for cations of opposite handedness linked to the same electron spin, and vice versa (see Fig.\ref{fig:idea_of_locking}).

\begin{figure}[t!]
\centering
\includegraphics[width=\textwidth]{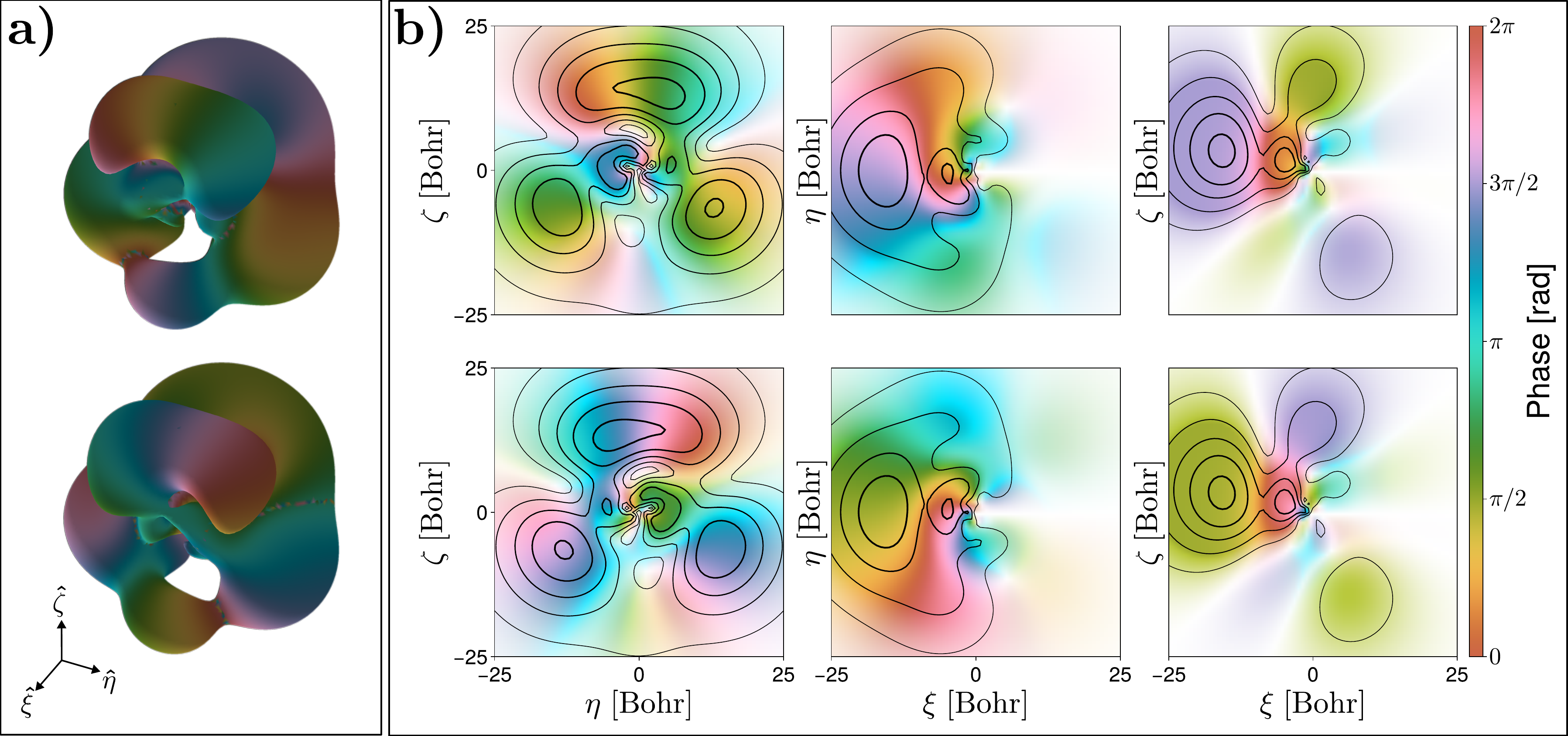}
\caption{Comparison of the (a) isosurface and (b) contour plots of the chiral electronic states $| \psi_{-1,\frac{1}{2}}^{+}\rangle$ and $| \psi_{-1,\frac{1}{2}}^{-}\rangle$ shown in the top and bottom row (in the molecular frame), respectively, and colored according to its phase. The molecular $\{x,y,z\}$ axes are labeled as $\{\xi,\eta,\zeta\}$, respectively. The isosurface is set at $|\psi_{-1,\frac{1}{2}}^{\pm}|=3.2\times10^{-3} \text{Bohr}^{3/2}$. The contour plots are cuts on the $\xi=0$, $\zeta=0$, and $\eta=0$ planes. Thicker contour lines and darker shading correspond to higher values of the density. It can be seen that the probability density of the states $| \psi_{-1,\frac{1}{2}}^{\pm}\rangle$ are mirror images of each other.
%, and the probability current rotates in opposite directions. 
} 
\label{fig:isosurface}
\end{figure}

To quantify this effect, we construct spin-resolved chiral electronic
states in Ar, rigorously incorporating electronic chirality and
spin--orbit coupling. Inspired by analogous chiral hydrogenic states
\cite{ordonez2019propensity}, the target states combine excited-state
orbitals as:
\begin{align}
| \psi_{m,\mu} \rangle^{\pm} = \dfrac{1}{\sqrt{2}} \left( |4p_{m},\mu\rangle \pm i |4d_{m},\mu\rangle \right).
\label{eq:c_state}
\end{align}
To realize such states in the multielectron Ar atom, we construct coherent
superpositions of its \emph{ab initio} spin--orbit-coupled excited states
using the optimization procedure detailed in
Appendix~\ref{app:chiral-argon}. A unitary transformation first
approximately disentangles the excited electron from the ionic degrees of
freedom, after which the resulting states are combined to reproduce the
desired chiral electronic superposition while maximizing the overlap of
the accompanying ionic states. The constructed states retain more than
$80\%$ weight in the targeted $n\ell m_\ell s m_s$ spin-orbitals. Unlike hydrogen, the multielectron core potential in argon breaks inversion symmetry, lowering the symmetry of the resulting states (Fig.\ref{fig:isosurface}). Consequently, synthetic chirality in argon is stabilized by electron correlations making this system  ideal for exploring spin-chirality coupling computationally within a fully consistent approach.

The synthetic chiral argon calculations employ \textit{ab initio} multichannel photoionization framework, with spin--orbit coupling incorporated directly into the electronic Hamiltonian. The electronic structure is described at the two-component configuration-interaction-singles level using relativistic effective-core potentials derived from multiconfigurational Dirac--Fock calculations, while the continuum amplitudes are obtained using the multichannel infinite-time surface-flux method with Coulomb asymptotics. The approach has been benchmarked against fully relativistic calculations and experimental atomic energies and reproduces established spin-resolved photoionization phenomena \cite{Carlstroem2024spinpolspectral,carlstrom2022general1,carlstrom2022general2}. Thus, while synthetic chiral argon provides a deliberately minimal system, its electronic structure, spin--orbit coupling, and photoionization dynamics are treated quantitatively from first principles. Argon provides a particularly stringent test: its chirality is purely electronic, its atomic spin--orbit interaction is weak (1\% of the ionization potential), and an atomic system excludes non-local spin--orbit interactions associated with neighboring atomic centers. We further average over all initial spin states, all molecular orientations, all degenerate states of the ion and all linear-polarization directions to realize isotropic illumination. Nevertheless, a pronounced spin--orientation correlation survives. Thus, while the system is deliberately minimal, the calculation itself is quantitative and \textit{ab initio}; establishing analytically and numerically that the correlation survives this extensive averaging is one of the central results of this work.

\begin{figure}[t!]
\centering
\includegraphics[width=\textwidth]{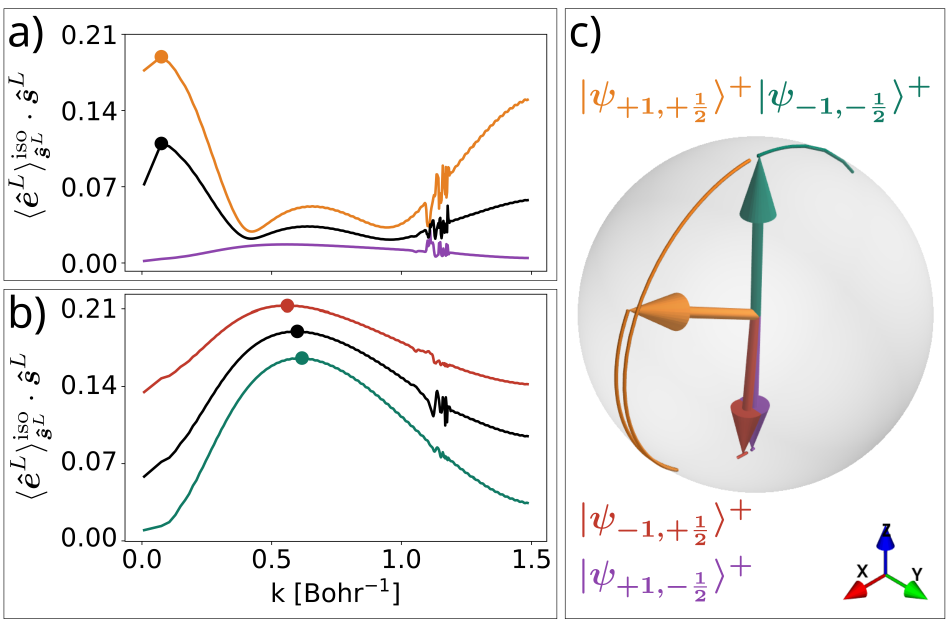}
\caption{Enantio-sensitive spin-orientation locking under isotropic illumination of randomly oriented electronic states. (a) Degree of orientation for chiral states with $m=1$, $\mu=\pm\frac{1}{2}$ (orange and violet correspondingly), and averaged over spin orientation in initial state (black). The dots mark the corresponding maximum degree of orientation $\langle \unitvec{e}\cdot\unitvec{s} \rangle^{\text{iso}}$. (b) Degree of orientation for chiral states with $m=-1$, $\mu=\pm\frac{1}{2}$ (red and green correspondingly) and averaged (black). The rapidly oscillating behavior at higher values of \(k\) are due to the Fano resonances, leading up to the ionization threshold for the 3s electrons \cite{Samson2002,Carlstroem2024spinpolspectral}. Dots indicate the maximal orientation vs $k$ for each initial state/ensemble. (c) The Bloch pseudovector $\fullvec{S}^M$ (internal directional bias) in the molecular frame for the chiral argon states. $\fullvec{S}^M$ changes its direction in space as a function of the photoelectron momentum $k$. Trajectories traced by this vector are shown for  $0<k<1.0 \text{ Bohr}^{-1}$.
}
\label{fig:spin_orient_lock_iso}
\end{figure}

Figure~\ref{fig:spin_orient_lock_iso} shows the degree of
spin--orientation locking for different chiral states in our synthetic
chiral system under isotropic illumination as a function of the
photoelectron momentum $k=\sqrt{2E}$. We quantify spin–orientation locking by the mean molecular orientation projected onto the selected spin-detection axis:
\begin{equation}
\mathcal O_{\hat s}
\equiv
 \langle \unitvec{e}^L \rangle_{\unitvec{s}^L}
^{\mathrm{iso}}
\cdot\unitvec{s}^{L},
\end{equation}
which ranges from zero for an unoriented ensemble to unity for perfect
orientation along the selected spin axis. For the molecular direction
of maximal correlation ($\cos\theta=1$), its magnitude is determined
directly by the Bloch pseudovector,
\begin{equation}
\mathcal O_{\hat s}^{\mathrm{max}}
=
\frac{\nu}{3}|\mathbf S^{M}|.
\end{equation}

The black curves in Fig.~\ref{fig:spin_orient_lock_iso}(a,b) show
photoionization from initially spin-unpolarized states, represented by
incoherent mixtures of the chiral states with $m=1$ and
$m_s=\pm1/2$ in panel (a), and $m=-1$ and $m_s=\pm1/2$ in panel (b).
Remarkably, pronounced spin--orientation locking persists even in this
minimal chiral system: we obtain $\mathcal O_{\hat s}^{\rm max}\simeq0.19$ for
$m=1$ and $\mathcal O_{\hat s}^{\rm max}\simeq0.22$ for $m=-1$.

For an intuitive estimate, we map these first moments onto the simple
head--tail angular-distribution model described in
Methods~\ref{subsec:estimate}. Within this approximate model,
$\mathcal O_{\hat s}^{\rm max}\simeq0.19$ ($0.22$) corresponds to approximately
$64\%$ ($66\%$) of the ionized molecules having a positive projection
of $\unitvec{e}^{L}$ onto the selected spin axis, compared with
$36\%$ ($34\%$) having a negative projection. These percentages provide
an intuitive, model-dependent interpretation of the orientation; the
directly calculated quantity $\mathcal O_{\hat s}$ is used as the
quantitative measure of spin--orientation locking.

The Bloch vectors for $m=1, m_s=\pm1/2$ are almost orthogonal to each other (see orange and violet arrows in Fig.\ref{fig:spin_orient_lock_iso} (c)), while for the state $m=+1,m_s=1/2$ the Bloch vector is orthogonal to the chosen spin quantization axis $\unitvec{\zeta}^M$. The latter signifies strong contribution of coherences to the longitudinal spin-orientation locking arising during photoionization. The role of coherences in longitudinal spin polarization has not been identified before, because the latter is usually identified as the z-component of the spin expectation value in the final state, which only reflects the difference in populations of the 
spin-up and spin-down final states at the detector. Figure \ref{fig:spin_orient_lock_iso}(c) also demonstrates that the direction of the Bloch vector changes as a function of the photoelectron energy: the tip of the unit vector representing the directions of Bloch pseudovector vs photoelectron energy traces a curve on the unit sphere.

\begin{figure}[t!]
\centering\includegraphics[width=\textwidth]{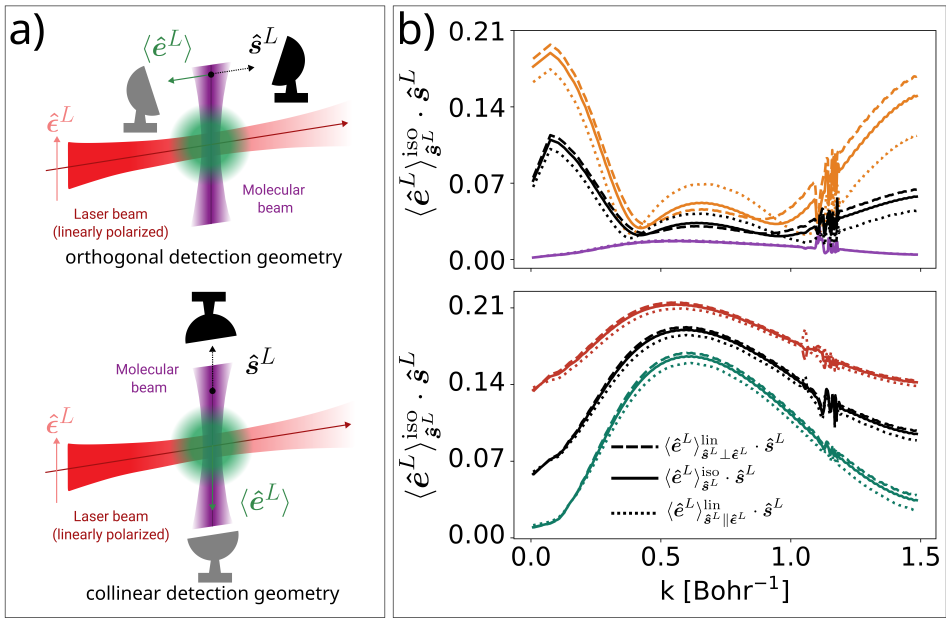}
\caption{ Enantio-sensitive spin-orientation locking resulting from illumination of randomly oriented electronic states by linearly polarized fields. (a) Proposed scheme for the orthogonal and collinear detection geometries, Eq. \eqref{eq:V-ave-lin}. (b) Comparison of the degree of orientation for isotropic illumination (solid line) and linearly polarized light with orthogonal (dashed line) and collinear (dotted line) detection geometries for the same chiral states as that of Fig. \ref{fig:spin_orient_lock_iso}.
}
\label{fig:ptype-Vave-lin}
\end{figure}

An additional directional bias introduced by the laser field may enhance the enantio-sensitive spin-orientation locking, e.g., polarization direction induces an additional directional bias quantified by the vector
\begin{align}
\fullvec{S'}^{M}=\frac{1}{S_0} \text{Re}\left[ \sum_{I,\mu, \nu} \int d\Theta_{k}^{M} \left(\fullvec{D}_{I,\fullvec{k}^{M},\mu}^{M*}\cdot\unitvec{\sigma}_{\nu,\mu}^{M}\right)\fullvec{D}_{I,\fullvec{k}^{M},\nu}^{M} \right]. 
\end{align}
For linearly polarized light, we obtain: 
\begin{align}
 \langle \unitvec{e}^L \rangle_{\unitvec{s}^L}^{\text{lin}} = \dfrac{1}{5} 
\begin{cases}
[(2\fullvec{S}^M - \fullvec{S'}^M) \cdot \unitvec{e}^M] \unitvec{s}^L \quad &,\quad \unitvec{s}^L \perp \unitvec{\epsilon}^L \\
[(\fullvec{S}^M + 2\fullvec{S'}^M) \cdot \unitvec{e}^M] \unitvec{s}^L  \quad &,\quad \unitvec{s}^L \parallel \unitvec{\epsilon}^L.
\end{cases}
\label{eq:V-ave-lin}
\end{align}
This leads to two possible detection geometries, in which the photoelectron spin can be detected either orthogonal ($\unitvec{s}^L\perp\unitvec{\epsilon}^L$) or parallel ($\unitvec{s}^L\parallel\unitvec{\epsilon}^L$) to the light field polarization vector $\unitvec{\epsilon}^L$, see Fig. \ref{fig:ptype-Vave-lin}(a). The axis of molecular orientation $\unitvec{e}_{\parallel}^M$ will then depend on the detection geometry, e.g., $(2\fullvec{S}^M - \fullvec{S'}^M) \equiv \nu |2\fullvec{S}^M - \fullvec{S'}^M| \unitvec{e}_{\parallel}^M $ for the orthogonal detection geometry. 
Figure \ref{fig:ptype-Vave-lin}(b) shows that for the chiral states $m=-1$ the orthogonal (parallel) detection geometry generally provides a slight enhancement (reduction). However, the chiral state $m=1,m_s=1$ (orange curve) shows that both detection geometries can provide enhancement within certain regimes of the photoelectron energy.

\section{Discussion}

\subsection{Spin--orientation correlations and the molecular compass}
\label{subsec:bloch_ciss}

One of the central results of our microscopic calculation is the emergence
of correlations between molecular orientation and the detected
photoelectron spin in a completely isotropic setting. The microscopic origin of the spin--orientation correlation lies in the
photoionization amplitude itself. The electric-dipole interaction,
$\fullvec{D}^{M}_{I,\fullvec{k}^{M},\mu}\cdot \fullvec{E}^{M}_{p}(\rho)$, makes the ionization probability sensitive to molecular orientation, while spin--orbit coupling renders the photoionization dipoles spin dependent. The spin-resolved
ionization rate defines two complementary conditioned
observables: (i) the mean spin direction for a selected molecular orientation:
\begin{equation}
\langle \unitvec{s}^L \rangle_{\rho} = \dfrac{1}{N} \int d\Theta_p \int d\Theta_s^L \int d\Theta_k^L \, \unitvec{s}^L W^{L}(\hat{\mathbf k}^{L},\hat{\mathbf s}^{L},\rho),
\label{eq:parallel_spin}
\end{equation}
and, (ii) the mean molecular orientation for a selected spin projection along $\unitvec{s}^L$:
\begin{equation}
\langle \unitvec{e}^L \rangle_{\unitvec{s}^L} = \dfrac{1}{N} \int d\rho \int d\Theta_p \int d\Theta_k^L \, \unitvec{e}^L W^{L}(\hat{\mathbf k}^{L},\hat{\mathbf s}^{L},\rho)
\label{eq:parallel_orientation}
\end{equation}
Equation \eqref{eq:parallel_spin} defines a fixed molecular orientation and averages the spin-detection axis  $\unitvec{s}^L$, with each axis weighted by the probability of detecting the positive spin projection along that direction; a non-zero first moment therefore directly signifies a net photoelectron spin polarization and determines its direction. Equation \eqref{eq:parallel_orientation} defines a fixed spin-detection axis and the molecular orientation is averaged (see Methods \ref{subsec:correlations}).
The two conditional averages are generally normalized by the corresponding
conditional yields and are typically not equal to each other. In the present isotropic setting, however, these
normalizations are common  $N=S_0|E_\omega^L|^2/6$. Indeed, isotropic illumination makes the spin-integrated ionization rate
independent of the molecular orientation $\rho$, while the absence of net
spin polarization makes the orientation-integrated rate independent of the
selected spin-detection axis $\hat{\mathbf s}^L$. With the normalized
measures $\int d\rho=1$ and $\int d\Theta_s^L=1$, both conditional
denominators are therefore equal to $N$.

The correlation between molecular orientation and photoelectron spin is therefore characterized by the corresponding second moment of the full joint distribution, i.e., 
\begin{equation}
C_{ij}^{L} (\unitvec{e}^M) \equiv
\left\langle e_i^{L}s_j^{L}\right\rangle
= \dfrac{1}{N} \int d\rho \int d\Theta_p \int d\Theta_s^L \int d\Theta_k^L \, e_i^{L}s_j^{L} W^{L}(\hat{\mathbf k}^{L},\hat{\mathbf s}^{L},\rho). 
\label{eq:C_joint}
\end{equation}
To expose the molecular-frame structure underlying the ensemble correlation
$C_{ij}^{L}(\hat{\mathbf e}^{M})$, we introduce its orientation-resolved
counterpart,
\begin{equation}
C_{ij}^{L} (\unitvec{e}^M) \equiv \langle G_{ij}^{L}(\rho;\unitvec{e}^M) \rangle = \int d\rho \, G_{ij}^{L}(\rho;\unitvec{e}^M)
\end{equation}
\begin{equation}
G_{ij}^{L}(\rho;\unitvec{e}^M) = \dfrac{1}{N}\int d\Theta_p \int d\Theta_s^L \int d\Theta_k^L \, e_i^{L}s_j^{L} W^{L}(\hat{\mathbf k}^{L},\hat{\mathbf s}^{L},\rho) = e_i^L \langle s_j^L \rangle_\rho 
\label{eq:G_joint_moment}
\end{equation}
The orientation resolved correlation tensor $\mathbf G^{L}$ is not introduced phenomenologically, but follows
directly as a mixed moment of the microscopic joint distribution. Since $\unitvec{e}^L$ is a unit vector, Eq. \eqref{eq:G_joint_moment} immediately gives
\begin{equation}
%\boxed{
\langle \unitvec{s}^L \rangle_\rho
=
\left[
\mathbf G^{L}
\left(
\rho;\hat{\mathbf e}^{M}
\right)
\right]^T
\hat{\mathbf e}^{L}.
%}
\label{eq:spin_from_G}
\end{equation}

To see that the $\langle \unitvec{e}^L \rangle_{\unitvec{s}^L}$ also arises from the same correlation tensor $G_{ij}$, it is helpful to represent the joint distribution as:
\begin{equation}
 W^{L}(\unitvec{k}^{L},\unitvec{s}^{L},\rho)
 =
\frac{1}{2}
\left(
 W_{0}^{L}
 +
 s_j^{L}W_{S,j}^{L}
 \right),
 \label{eq:split_rate}
 \end{equation}
where the second term expresses spin-dependent part of the joint ionization rate in Eq. \eqref{eq:ion_rate} in terms of the scalar product with:
\begin{equation}
\mathcal{W}_{S,j}^L = \dfrac{1}{S_0|E_\omega^L|^2}\int d\Theta_p \int d\Theta_k^L \, \text{Tr} \left[ \sum_I \left( \fullvec{D}_{I,\fullvec{k}^{M}}^{L} \cdot \fullvec{E}_p^{L}\right) \unitvec{\sigma} \left( \fullvec{D}_{I,\fullvec{k}^{M}}^{L*} \cdot \fullvec{E}_p^{L*}\right) \right]_j. 
\label{eq:W_Sj}
\end{equation}
wherein the trace is performed in spin space and the index $j$ denotes the $j^{th}-$component of the vector quantity $\text{Tr}(\dots)$. Using Eqs.\eqref{eq:ion_rate},\eqref{eq:parallel_spin}, \eqref{eq:G_joint_moment}, and \eqref{eq:split_rate}, we obtain 
\begin{equation}
\langle s_i^L \rangle_\rho = \dfrac{1}{3} \mathcal{W}_{i}^L(\rho) 
% \label{eq:spin_fixed_rho}
\quad,\quad 
G_{ij}^{L}(\rho;\unitvec{e}^M) = \dfrac{1}{3} e_i^L \mathcal{W}_{j}^L(\rho). 
\label{eq:eWs_G}
\end{equation}
Finally, using Eq. \eqref{eq:eWs_G} together with isotropy of the
spin-independent contribution and the absence of net spin polarization, we obtain the connection between the spin-conditioned orientational moment and the correlation tensor: 
\begin{equation}
\langle e_i^L \rangle_{\unitvec{s}^L} = 3 \left( \int d\rho\, e_i^L \mathcal{W}_{j}^L(\rho) \right) s_j^L = 3 \langle G_{ij}^{L}(\rho;\unitvec{e}^M) \rangle \, s_j^L.
\label{eq:micro-G}
\end{equation}

Having connected the correlation tensor to both conditioned moments we can use our earlier microscopic results (Eq. \eqref{eq:microscopic}) to precisely determine this tensor. 
Indeed, the orientation-resolved correlation tensor has the exact microscopic form as that of Eq. \eqref{eq:G_joint_moment}, i.e., 
\begin{equation}
\mathbf G^{M}(\rho;\unitvec{e}^{M}) = \unitvec{e}^{M}\otimes \langle \unitvec{s}^{M}\rangle_\rho,
\label{eq:G_exact_M}
\end{equation}
which is an outer product of the molecular orientation $\unitvec{e}^M$ and the mean spin direction for a selected molecular orientation, and has rank one for a non-vanishing correlation. Since the molecular orientation $\hat{\mathbf e}$ is a polar vector, whereas the spin-detection direction $\hat{\mathbf s}$ is axial and $\mathbf G^{M}$ is therefore a pseudotensor. Since its first direction is fixed by
$\unitvec{e}^{M}$, then its most general rank-one representation is
\begin{equation}
\mathbf G^{M}(\rho;\unitvec{e}^{M}) = g_{\parallel}\, \unitvec{e}^{M} (\unitvec{b}^{M})^T,
\qquad |\unitvec{b}^{M}|=1,
\label{eq:G_rank_one}
\end{equation}
where the spin-side direction unit polar vector $\unitvec{b}^{M}$ and pseudoscalar $g_{\parallel}$ are initially
undetermined.   The isotropic rotational average of a molecular-frame tensor obeys $\langle
R_{\rho}\mathbf G^{M}R_{\rho}^{T}\rangle=\frac{1}{3}\text{Tr}(\boldsymbol{G}^M\mathbb{I})$, therefore, 
\begin{equation}
\langle\mathbf G^{L} (\rho;\unitvec{e}^{M}) \rangle = \dfrac{g_{\parallel}}{3} \left( \unitvec{e}^{M}\cdot\unitvec{b}^{M} \right) \mathbb I,
\label{eq:G_rank_one_average}
\end{equation}
together with the microscopic calculation (Eq. \eqref{eq:microscopic}) and Eq.\eqref{eq:micro-G} gives the chain of equations that connects the microscopic result on the left with the result expressed via correlation tensor on the right:
\begin{equation}
\langle \unitvec{e}^L \rangle_{\unitvec{s}^L}=\dfrac{\nu}{3}|\fullvec{S}^M| \left( \unitvec{e}_{\parallel}^M \cdot \unitvec{e}^M \right)\unitvec{s}^L= 3\langle\mathbf G^{L} (\rho;\unitvec{e}^{M}) \rangle \unitvec{s}^L=g_{\parallel}\left( \unitvec{e}^{M}\cdot\unitvec{b}^{M} \right)\unitvec{s}^L.%\dfrac{2}{3} \left( \fullvec{S}^{M}\cdot\unitvec{e}^{M} \right)\mathbb I .
\label{eq:G_micro_average}
\end{equation}
Since  left and right parts of these equations must be equal for any $\unitvec{e}^{M}$,
we find that the microscopic result uniquely identifies the spin-side correlation
axis with the photoionization Bloch-vector axis,
\begin{equation}
\unitvec{b}^{M}=\unitvec{e}_{\parallel}^{M},
\qquad
g_{\parallel}
=
\frac{\nu}{3}|\fullvec{S}^{M}|,
\end{equation}
and hence
\begin{equation}
%\boxed{
\mathbf G^{M}(\rho;\unitvec{e}^{M})
=
g_{\parallel}\,
\unitvec{e}^{M}
(\unitvec{e}_{\parallel}^M)^T.
%}
\label{eq:G_final}
\end{equation}
The complementary orientation-conditioned response now follows directly
from Eq. \eqref{eq:G_exact_M},
\begin{equation}
%\boxed{
\langle \unitvec{s}^M \rangle_\rho 
=
\left[\mathbf G^{M}( \unitvec{e}^{M})\right]^{T}
\unitvec{e}^{M}
=
g_{\parallel}\unitvec{e}_{\parallel}^{M}.
%}
\label{eq:orientation_cond}
\end{equation}
Thus, for a molecule of fixed orientation, the mean photoelectron spin is
locked to the molecular photoionization Bloch-vector axis. Its magnitude
is set by $|\fullvec{S}^{M}|$, while its enantio-sensitive sign is carried
by the pseudoscalar $g_{\parallel}$.
For an orientation-conditioned measurement, exchange of enantiomers
reverses the molecular pseudoscalar,
$g_{\parallel}^{(R)}=-g_{\parallel}^{(L)}$, and therefore reverses the
orientation-conditioned spin response  for the same conditioned molecular
orientation,
\begin{equation}
\langle \unitvec{s} \rangle_\rho ^{(R)}
=
-
\langle \unitvec{s} \rangle_\rho^{(L)}.
\end{equation}
Thus the pseudoscalar character of $g_{\parallel}$ provides the
enantio-sensitive sign of the spin response, whereas
$\unitvec{e}^{M}\parallel\hat{\mathbf S}^{L}$ specifies its direction.

%%%%%%%%%%%%%%%%%%%%%%%
\section{Conclusions and Outlook} 
We established the two complementary manifestations of the same microscopic
spin--orientation correlation: (i) the orientation conditioned measurement of mean spin direction:
\begin{equation}
\langle \unitvec{s}^L \rangle _\rho = g_{\parallel} \unitvec{e}_\parallel^L,
\label{eq:s_fin}
\end{equation}
and, (ii) the spin conditioned measurement of mean molecular orientation:
\begin{equation}
\langle \unitvec{e}^L \rangle _{\unitvec{s}^L} = g_{\parallel} \cos\theta \unitvec{s}^L.
\end{equation}
Selecting either the molecular orientation or photoelectron spin thereby exposes complementary first moments of the same joint distribution determined by the spin-dependent ionization rate. The direction of the photoionization Bloch vector defines the molecular axis along which this correlation is maximal, while its magnitude determines the strength of the corresponding molecular compass. Orientation-conditioned measurement corresponds to CISS, while spin-conditioned measurement corresponds to spin-orientation locking predicted in this work.

The absence of an additional angular factor in
Eq.~\eqref{eq:s_fin} is important. The factor $\cos\theta = \unitvec{e}^M \cdot \unitvec{e}_\parallel^M$ appearing in the spin-conditioned orientation $\langle \unitvec{e}^L \rangle_{\unitvec{s}^L}^{\text{iso}} = g_{\parallel}\cos\theta\, \unitvec{s}^{L}$ quantifies how efficiently the chosen molecular vector $\unitvec{e}^{M}$ reads out the intrinsic compass direction $\unitvec{e}_\parallel^M$. By contrast,
$\langle\unitvec{s}^{M}\rangle_\rho$ is conditioned on the full
molecular orientation $\rho$ and is therefore independent of which
molecular vector is used to represent that orientation. For example, if
$\unitvec{e}^{M}\perp\unitvec{e}_{\parallel}^{M}$,
\begin{equation}
\langle \unitvec{e}^L \rangle_{\unitvec{s}^L}
=0,
\qquad
\langle
\unitvec{s}^{M}
\rangle_\rho
=
g_{\parallel}\unitvec{e}_{\parallel}^{M}\neq0.
\end{equation}
There is no contradiction: a molecule of known full orientation retains
its spin response along the intrinsic compass direction, whereas a
molecular vector perpendicular to this direction carries no net
spin-conditioned orientational signal after isotropic averaging.

Although the Bloch pseudovector can exist in achiral molecules, parity symmetry forbids any spin–orientation locking.
In an isotropically illuminated, randomly oriented ensemble such as, e.g., HCl, mirror reflection leaves both the molecular distribution and the light field unchanged.
The laboratory spin-detection axis $\hat{\mathbf s}^L$, being a pseudovector, also remains invariant under reflection.
If a correlation between spin and molecular orientation were present—for instance, H–Cl cations preferentially associated with spin-up electrons—the mirrored configuration would produce the opposite cation orientation (Cl–H) for the same spin direction.
Because the two mirror-related outcomes are equally probable, their contributions cancel in the ensemble average, and no net spin–orientation correlation can survive in achiral molecules.

Our analysis shows that quantum correlations between the photoelectron spin and the molecular skeleton orientation are unique to chiral molecules. Importantly, they survive averaging over random molecular orientations, with the coherence between the spin-up and spin-down final components of the photoelectron recorded in the Bloch pseudovector resulting from the reduced density matrix. These correlations underlie the main microscopic mechanism of chirality induced spin selectivity. Remarkably, we show that the strength of the spin-orientation correlations, the magnitude of the CISS effect in photoionization and the spin-orientation locking are all characterized by the direction and the magnitude of the Bloch vector. We identify the photoionization Bloch vector as a chiral molecular compass mediating spin-chirality coupling and offering significant control over the orientation of enantiomers in space. Substituting the photoionization dipoles by the photoexcitation dipoles, our results translate to photoexcitation and extend the concept of enantiosenstive molecular compass to the photoexcitation processes. Thus the chiral molecular compass also gives rise to enantio-sensitive spin dynamics in photoexcited oriented molecules. Our approach highlights that both the direction and magnitude of the chiral molecular compass depend sensitively on photon energy, leading to either strong or weak spin-orientation correlations  and associated spin polarization or orientation phenomena within the same chiral medium. We expect that the enantio-sensitive molecular compass is also relevant for understanding  electron spin polarization in scattering from gas-phase chiral molecules \cite{campbell1987electron,mayer1995experimental} - the phenomenon that predated the formulation of the CISS effect.

\section{Methods}
\label{sec:methods}

\subsection{Bloch vector associated with spin-resolved photoionization}
\label{subsec:bloch}

Using perturbation theory, the full spinor electron wave-function at the end of the ionizing pulse is: 
\begin{subequations}
\label{eqn:pt-wavefunction}
\begin{equation}		|\psi\rangle=|\psi_{o}\rangle+
\sum_{I,\mu^M}
%\int\mathrm{d}k
\int\mathrm{d}\Theta_k^M\, c_{I,\fullvec{k}^M,\mu^M}
|I\Psi_{I,\fullvec{k}^M,\mu^M}^{(-)}\rangle , %|\chi_{\mu^M}\rangle ,
\label{eq:psi}
\end{equation}
\begin{equation}
c_{I,\fullvec{k}^M,\mu^M} =
i\left( \fullvec{D}_{I,\fullvec{k}^M,\mu^M}^L\cdot\fullvec{E}^L \right),
\end{equation}
\end{subequations}
where, $|\Psi_{I,\fullvec{k},{\mu}}^{(-)}\rangle$ are fully spin-orbit coupled continuum states, which are the two components of a spinor-valued scattering solution with opposite projections of spin (${\mu}=\pm\frac{1}{2}$) onto the molecular z-axis
$\unitvec{\zeta}$
subject to the orthogonality condition $\langle I\Psi_{I,\fullvec{k},\mu}^{(-)}|I'\Psi_{I',\fullvec{k'},\nu}^{(-)}\rangle=\delta_{I,I'}\delta(\fullvec{k}-\fullvec{k'})\delta_{\mu,\nu}$. The ion channel is denoted by $I$, while $\fullvec{D}_{I,\fullvec{k}^M,\mu^M}^L=\langle I\Psi_{I,\fullvec{k}^M,\mu^M}^{(-)}|\fullvec{d}^L|\psi_0\rangle$ 
is the transition dipole matrix element, and $\fullvec{E}^L$ is the light field.

We now consider the full density operator describing the spin and spatial degrees of freedom for ionization into the final photoelectron momentum $k=\sqrt{2E}$ and (degenerate) ion channels $I$ (fixed by the energy conservation):
\begin{align}
\hat{\varrho}^{M} = \sum_{I,I'}\int d\Theta_k^M \int d\Theta_{k'}^M\sum_{\mu^M,\nu^M} c_{I,\fullvec{k}^M,\mu^M}c_{I',\fullvec{k'}^M,\nu^M}^{*}|I\Psi_{I,\fullvec{k'}^M,\mu^M}\rangle\langle\Psi_{I,\fullvec{k}^M,\nu^M}I'|.
\label{eq:varrho}
\end{align}
To obtain the reduced spin-space density matrix, we perform a partial trace over the spatial continuum states and (degenerate) ionization channels:
\begin{align}
{\varrho}_{\mu,\nu}^{M} =& \mathrm{Tr}_{\text{(spatial+channels)}}(\hat{\varrho}^{M}) = \sum_{I} \int d\Theta_k^M \langle I\Psi_{I,\fullvec{k}^M,\mu^M}|\hat{\varrho}^{M}|I\Psi_{I,\fullvec{k}^M,\nu^M}\rangle.
\label{eq:trace}
\end{align}
Using Eqs. \eqref{eq:varrho} and \eqref{eq:trace}, we obtain:
\begin{align}
{\varrho}_{\mu,\nu}^{M}
&= \sum_{I}\int d\Theta_k^M c_{I,\fullvec{k}^M,\mu^M}c_{I,\fullvec{k}^M,\nu^M}^*,
\label{eq:matden}
\end{align}
which follows from the orthogonality condition. 

A possible way of introducing the Bloch vector describing spin orientation of a two-level spin $1/2$ system  defined as $\fullvec{P} \equiv \mathrm{Tr}(\varrho^{M}\unitvec{\sigma}^M)$ is to use $\varrho^{M}$ with the matrix elements given by Eq. \eqref{eq:matden}. In this case, the Bloch vector would depend both on the properties of the laser field as well as the molecular geometry, which does not characterize an \emph{intrinsic} enantio-sensitive spin orientation locking that solely relies on the molecular geometry.
%as it follows from  first principles derivation of spin-orientation locking exposed in the last subsection of Methods.
We now introduce the photoionization Bloch vector $\fullvec{S}^{M}$ by averaging $\varrho_{\mu,\nu}^M$ over random molecular orientations, i.e., 
\begin{subequations}
\begin{align}
\int d\rho \, \varrho_{\mu,\nu}^M =& \sum_{I} \int d\Theta_k^M \left[ \int d\rho \left( \fullvec{D}_{I,\fullvec{k}^M,\mu^M}^L \cdot \fullvec{E}^L \right) \left( \fullvec{D}_{I,\fullvec{k}^M,\nu^M}^{L*} \cdot \fullvec{E}^{L*} \right) \right] 
= \dfrac{1}{3} |\fullvec{E}^L|^2 \tilde{\varrho}_{\mu,\nu}^M 
\end{align}
\begin{equation}
\tilde{\varrho}^{M}_{\mu,\nu} 
= \sum_{I}\int d\Theta_{k}^{M}\left( \fullvec{D}_{I,\fullvec{k}^{M},\mu}^{M}\cdot\fullvec{D}_{I,\fullvec{k}^{M},\nu}^{M*} \right).
\label{eq:density_matrix_defM}
\end{equation}
\end{subequations}
Normalizing $\tilde{\varrho}^{M}_{\mu,\nu} $ in a standard way to its trace $\mathrm{Tr}[\tilde\varrho^{M}]=S_0$, where 
$S_0$ is proportional to the total ionization rate: 
\begin{align}
S_0 =\sum_{I} \left[  \int d\Theta_k^M\left(  \left| \fullvec{D}_{I,\fullvec{k}^M,\frac{1}{2}}^M \right|^2 +  \left| \fullvec{D}_{I,\fullvec{k}^M,-\frac{1}{2}}^M \right|^2 \right) \right] 
\end{align}
we can write in the following equivalent form: 
\begin{equation}
\frac{\tilde\varrho^{M}}{\mathrm{Tr}[\tilde\varrho^{M}]}
= \frac{1}{2}\left(\,\mathbb{I} + \fullvec{S}^{M}\cdot\unitvec{\sigma}^M\right).
\label{eq:density_matrix_decompositionM}
\end{equation}
Here, $\fullvec{S}^{M} $ is the photo-ionization Bloch pseudovector in real space
\begin{equation}
\fullvec{S}^{M} 
= \frac{1}{S_0}\mathrm{Tr}\left(\tilde{\varrho}^{M}\unitvec{\sigma}^M\right),
\label{eq:spin_polarization_vectorM}
\end{equation}
and is an intrinsic molecular property invariant under rotations of the spin quantization axis provided that these rotations are applied consistently to both the density matrix $\tilde{\varrho}^M$ and the spin operator $\unitvec{\sigma}^M$. 

\subsection{Quantifying spin-orientation locking}
\label{subsec:quantify}

The spin-resolved ionization rate for a given orientation is obtained by projecting 
the full wavefunction Eq. \eqref{eqn:pt-wavefunction} onto the scattering and ionic states onto the spin detection axis $\unitvec{s}^L$ 
with energy $\mathcal{E}$, i.e., 
\begin{align}
	W^L(\unitvec{k}^L,\unitvec{s}^L,\rho) = \langle \psi | \mathsf{\hat{P}}_{\mathcal{E}} | \psi \rangle = \dfrac{1}{2}\sum_{I,\mu,\nu} \left( \fullvec{D}_{I,\fullvec{k}^{M},\mu}^{L*} \cdot \fullvec{E}^{L*}\right)\left( \fullvec{D}_{I,\fullvec{k}^{M},\nu}^{L} \cdot \fullvec{E}^{L}\right) \left( \delta_{\mu,\nu} + \unitvec{s}^L \cdot \unitvec{\sigma}_{\nu,\mu}^L \right) 
	\label{eq:methods_yield}
\end{align}
\begin{equation}
	\mathsf{\hat{P}}_{\mathcal{E}} = \sum_{I,\mu,\nu} 
	|I \Psi_{I,\vec{k}^M,\mu}^{(-)}\rangle \left[ \dfrac{\mathbb{I} +\unitvec{s}^L\cdot\unitvec{\sigma}^L}{2} \right]_{\mu,\nu}  \langle I \Psi_{I,\vec{k}^M,\nu}^{(-)}|, 
\end{equation}
where, $W^L(\unitvec{k}^L,\unitvec{s}^L,\rho)$ is the the momentum- and spin-resolved rate for a specific molecular orientation $\rho$,
$\unitvec{\sigma}$ is the vector of Pauli spin matrices, and we introduced the vector $\unitvec{\sigma}_{\mu,\nu}^L= \langle \chi_{\nu} | \unitvec{\sigma}^M | \chi_{\mu} \rangle$. %(see Appendix \ref{app:rate}).

Equation \eqref{eq:methods_yield} fixes the photoelectron momentum and spin-detection axis in the molecular and laboratory frame, respectively. By doing so, the transition dipole matrix element in the molecular frame $\fullvec{D}_{I,\fullvec{k}^M,\mu^M}^M$ will not have any argument that depends on the orientation $\rho$, and can therefore be trivially rotated into the lab frame: 
\begin{align}
\fullvec{D}_{I,\fullvec{k}^M,\mu^M}^L =& \langle I \Psi_{I,\fullvec{k}^M,\mu^M}^{(-)} | \fullvec{d}^L | \psi_o \rangle = \langle I \Psi_{I,\fullvec{k}^M,\mu^M}^{(-)}  | R_\rho \fullvec{d}^M | \psi_o \rangle = R_\rho \langle I \Psi_{I,\fullvec{k}^M,\mu^M}^{(-)}  | \fullvec{d}^M | \psi_o \rangle = R_\rho \fullvec{D}_{I,\fullvec{k}^M,\mu^M}^M. 
\end{align}
Meanwhile, by fixing the spin-detection axis in the lab frame, the spin projection operator $\hat{\mathsf{P}}_{\unitvec{s}}=\left(\mathbb{I}+\unitvec{s}^L\cdot\unitvec{\sigma}^L\right)/2$ essentially rotates the direction of photoelectron spin from the molecular to the laboratory frame, then projects it to the spin-detection axis $\unitvec{s}^L$:
\begin{align}
\unitvec{s}^L\cdot\unitvec{\sigma}_{\nu,\mu}^L = \unitvec{s}^L \cdot \bigg( R_\rho \unitvec{\sigma}_{\nu,\mu}^M \bigg).
\end{align} 
Thus, Eq. \eqref{eq:methods_yield} can be written as 
\begin{align}
W^L&(\unitvec{k}^L,\unitvec{s}^L,\rho) = \dfrac{1}{2}\sum_{I,\mu,\nu} \left[ \left( R_\rho\fullvec{D}_{I,\fullvec{k}^M,\mu}^{M*} \right)\cdot\fullvec{E}^{L*} \right]  \left[ \left( R_\rho\fullvec{D}_{I,\fullvec{k}^M,\nu}^{M} \right)\cdot\fullvec{E}^{L} \right] \left\{ \delta_{\mu,\nu} + \bigg[ \unitvec{s}^L \cdot \left( R_\rho \unitvec{\sigma}_{\nu,\mu}^M \right) \bigg] \right\}
\label{eq:methods_averaging_app_2}
\end{align}
The vectors that appear on the right-hand side of Eq. \eqref{eq:methods_averaging_app_2} can now be grouped into two sets: (i) vectors that are fixed in the molecular frame such as the dipole transition vectors $\fullvec{D}_{I,\fullvec{k}^M,\mu^M}^M$, photoelectron momentum $\fullvec{k}^M$, and photoelectron spin expectation value $\unitvec{\sigma}_{\nu,\mu}^M$, and (ii) vectors that are fixed in the laboratory frame such as spin detection axis $\unitvec{s}^L$ and the electric field $\fullvec{E}^L$. This will then allow us to use the technique in Ref. \cite{andrews1977three} in evaluating the orientation averaging $\int d\rho$ such that the resulting quantity can be expressed as $\sum_{ij}g_i M_{ij} f_j$, where, $g_i$ and $f_i$ are rotational invariants that are constructed from the two sets of vectors and $M_{ij}$ is the coupling between the two rotational invariants.

Let us now consider an arbitrary molecular unit vector $\unitvec{e}^{M}$ and evaluate its laboratory-frame expectation value
$\unitvec{e}^{L}(\rho)$ over the ionization probability conditioned on detecting the photoelectron along the laboratory-frame axis
$\unitvec{s}^{L}$:
\begin{subequations}
\begin{equation}
\langle \unitvec{e}^L \rangle_{\unitvec{s}^L}^{\text{iso}} \equiv \dfrac{1}{N} \int d\Theta_p \int d\rho \int d\Theta_k^L \, \unitvec{e}^L W^{L}(\hat{\mathbf k}^{L},\hat{\mathbf s}^{L},\rho) 
\end{equation}
\begin{equation}
N = \int d\Theta_p \int d\rho \int d\Theta_s^L \int d\Theta_k^L W^{L}(\hat{\mathbf k}^{L},\hat{\mathbf s}^{L},\rho)  \\ 
\end{equation}
\label{eq:ave}
\end{subequations}
Instead of a single direction that characterizes the direction of polarization, e.g., $\unitvec{x}^L$, we introduce the light field
\begin{align}
\fullvec{E}^L_p = E_\omega^L \left( \sin\theta_p\cos\varphi_p \unitvec{x}^L + \sin\theta_p\sin\varphi_p \unitvec{y}^L + \cos\theta_p \unitvec{z}^L \right)
\label{eq:E_iso2}
\end{align}
Combining Eqs. \eqref{eq:methods_averaging_app_2}-\eqref{eq:E_iso2}, and performing the necessary operations we obtain:
\begin{align}
N =& \dfrac{1}{6} \left[ \sum_{I,\mu} \int d\Theta_k^M \left| \fullvec{D}_{I,\fullvec{k}^{M},\nu}^{M} \right|^2 \right] |E_\omega^L|^2 = \dfrac{1}{6}S_0 |E_\omega^L|^2.
\end{align} 
which makes 
\begin{align}
\langle \unitvec{e}^L \rangle_{\unitvec{s}^L}^{\text{iso}} =&  \dfrac{6}{|E_\omega^L|^2 S_0} 
 \left \{ \int d\Theta_k^M \sum_{I,\mu,\nu}
\begin{bmatrix}
(  \fullvec{D}_{I,\fullvec{k}^{M},\mu}^{M*} \cdot \fullvec{D}_{I,\fullvec{k}^{M},\nu}^{M}) (\unitvec{\sigma}_{\nu,\mu}^M \cdot \unitvec{e}^M) \\ 
(\fullvec{D}_{I,\fullvec{k}^{M},\mu}^{M*} \cdot \unitvec{\sigma}_{\nu,\mu}^M)(\fullvec{D}_{I,\fullvec{k}^{M},\nu}^{M}\cdot\unitvec{e}^M) \\
(\fullvec{D}_{I,\fullvec{k}^{M},\mu}^{M*}\cdot\unitvec{e}^M)(\fullvec{D}_{I,\fullvec{k}^{M},\nu}^{M}\cdot\unitvec{\sigma}_{\nu,\mu}^M)
\end{bmatrix}^T
\right\} \nonumber \\
& \qquad \dfrac{1}{30} 
\begin{bmatrix}
    4 & -1 & -1 \\
    -1 & 4 & -1 \\
    -1 & -1 & 4
\end{bmatrix} \left\{ \int d\Theta_p
\begin{bmatrix}
|\fullvec{E}_p^L|^2 \unitvec{s}^L \\
(\unitvec{s}^L\cdot\fullvec{E}_p^L)\fullvec{E}_p^L  \\
(\unitvec{s}^L\cdot\fullvec{E}_p^L)\fullvec{E}_p^L
\end{bmatrix}
\right\}
=\dfrac{1}{3}(\fullvec{S}^M \cdot \unitvec{e}^M)\unitvec{s}^L.
\end{align}
Here, the vector of lab invariants become $\vec{\mathbb{F}}\propto [3 ,1,1]^T \unitvec{s}^L$ after integrating over all polarization directions $\Theta_p$, and that the coupling matrix $\mathbb{M}$ together with $\vec{\mathbb{F}}$ becomes proportional to $[1,0,0]^T$ such that only the first element of the molecular invariant $\vec{\mathbb{G}}$ is relevant.  

The product $\fullvec{S}^M\cdot\unitvec{e}^M$ is maximal when $\fullvec{S}^M\parallel\fullvec{e}^M$, thus, we can identify $\fullvec{S}^M$ with the molecular axis that becomes oriented.

A similar straightforward calculation for a light field polarized along $\unitvec{\epsilon}^L$ will yield 
\begin{align}
\langle \unitvec{e}^L \rangle_{\unitvec{s}^L}^{\text{lin}} = \dfrac{1}{5 S_0} \left[ \left( 2\fullvec{S}^M - \fullvec{S}'^M \right) \cdot \unitvec{e}^M  \right] \unitvec{s}^L - \dfrac{1}{5S_0} \left[ \left( \fullvec{S}^M - 3 \fullvec{S}'^M \right) \cdot \unitvec{e}^M  \right] ( \unitvec{s}^L \cdot \unitvec{\epsilon}^L ) \unitvec{\epsilon}^L. 
\label{eq:e_linear}
\end{align}
Equation \eqref{eq:e_linear} now presents two detection geometries: (i) orthogonal $\unitvec{s}^L\perp\unitvec{\epsilon}^L$, and (ii) collinear $\unitvec{s}^L\parallel\unitvec{\epsilon}^L$ which results to different axes of molecular orientation as shown in Eq. \eqref{eq:V-ave-lin}. Here, the additional vector $\fullvec{S}'^M$ has the form
\begin{align}
\fullvec{S'}^{M}= \frac{1}{S_0}\text{Re}\left[ \sum_{I,\mu,\nu} \int d\Theta_{k}^{M} \left(\fullvec{D}_{I,\fullvec{k}^{M},\mu}^{M*}\cdot\unitvec{\sigma}_{\nu,\mu}^{M}\right)\fullvec{D}_{I,\fullvec{k}^{M},\nu}^{M} \right],
\end{align}
which presents the directional bias induced by the well-defined direction of light polarization $\unitvec{\epsilon}^L$.

\subsection{Synthetic chiral matter}
\label{subsec:chiral-argon}

Quantification of spin-chirality coupling in photoionization requires a system where both electronic chirality and spin-orbit coupling are rigorously accounted. To this end, we construct chiral electronic densities in an  Ar atom corresponding to excitation into a chiral superposition of excited states, resolved on the excited electron spin. The eigenstates and associated spin-resolved photoionization dipole matrix elements of Ar atom were calculated using an atomic configuration-interaction singles treatment \cite{carlstrom2022general1,carlstrom2022general2,Carlstroem2024spinpolspectral}:
\begin{align}
\fullvec{D}_{I,\fullvec{k}^M,\mu^M}^M \equiv & 
\matrixel*{I\Psi_{I,\fullvec{k},{\mu^M}}^{(-)}}{\fullvec{d}^M}{\psi_o} = \bra*{\psi_{I,\fullvec{k},{\mu^M}}^{(-)}}\langle\chi_{\mu^M}|
[\hat{h}_S,\theta(r-r_s)] \sqrt{N}
\matrixel*{I}{(\hat{H}_0-\epsilon)^{-1}\fullvec{d}^M}{\psi_o},
\label{eqn:ionization-dipole-matrix-element}
\end{align}

In Eq. \eqref{eqn:ionization-dipole-matrix-element}, the volume integral in the matrix element is replaced with a surface integral over a sphere of radius \(r_s\) where the exact
continuum state is matched to the scattering state in the 
asymptotic region \cite{Carlstroem2024spinpolspectral} and a time integral which has been explicitly evaluated. The resolvent \((\hat{H}_0-\epsilon)^{-1}\) formally propagates the component of \(\fullvec{d}^M\ket{\psi_o}\) with energy \(\epsilon=E_I + k^2/2\) to infinite time, where \(E_I\) is the energy of the ion state \(I\), and the projection from the left by \(\sqrt{N}\bra{I}\) yields an energy-resolved Dyson orbital ($N$ is the number of electrons). The commutator of the scattering Hamiltonian \(\hat{h}_S\), i.e., the asymptotic Hamiltonian obeyed by the wave function and the scattering state in the region beyond the matching surface $r_s$  and the Heaviside function \(\theta(r-r_s)\) reduces the remaining one-electron volume integral to the surface integral \cite{Carlstroem2024spinpolspectral}.

Chiral spin resolved electronic states in Ar are inspired by a similar spinless chiral superposition in the hydrogen atom \cite{ordonez2019propensity}, and in the case of hydrogenic wave-functions could be represented as:
\begin{align}
| \psi_{m,\mu}^{\pm} \rangle =& \dfrac{1}{\sqrt{2}} \left( |4p_{m},\mu\rangle \pm |4d_{m},\mu\rangle  \right).
\label{eq:p_stateM}
\end{align}
Aiming to create analogues of such states in a multielectron system such as the argon atom, we devised an optimization procedure that yields the best approximation to these states (see Appendix \ref{app:chiral-argon} for details).

\subsection{Estimating the degree of orientation}
\label{subsec:estimate}

To provide an intuitive interpretation of the calculated orientational
first moment, we estimate the fractions of molecules whose orientation
vector has a positive (``head'') or negative (``tail'') projection onto
the selected spin-detection axis. This estimate is obtained by adopting
the approach of Ref.~\cite{ordonez2023geometric} to model the angular
distribution of oriented molecules.

We characterize the spin-conditioned orientation by
\begin{equation}
\mathcal O_{\hat s}
\equiv
 \langle \unitvec{e}^L \rangle_{\unitvec{s}^L}
^{\mathrm{iso}}
\cdot\unitvec{s}^{L}.
\label{eq:orientation_parameter}
\end{equation}
Without loss of generality, we choose
$\unitvec{s}^{L}\parallel\unitvec{z}^{L}$, such that
$\mathcal O_{\hat s}=\langle\cos\theta\rangle$, where $\theta$ is the
angle between the molecular orientation vector $\unitvec{e}^{L}$ and
$\unitvec{z}^{L}$. Following Ref.~\cite{ordonez2023geometric}, we model the corresponding angular distribution using
\begin{equation}
\Psi(\theta,\varphi)
=
a_0Y_{0,0}(\theta,\varphi)
+b_0Y_{1,0}(\theta,\varphi)
+\frac{b_-}{\sqrt{2}}
\left[
Y_{1,-1}(\theta,\varphi)-Y_{1,1}(\theta,\varphi)
\right]
+i\frac{b_+}{\sqrt{2}}
\left[
Y_{1,-1}(\theta,\varphi)+Y_{1,1}(\theta,\varphi)
\right],
\end{equation}
with
\begin{equation}
|a_0|^2+|b_0|^2+|b_-|^2+|b_+|^2=1.
\end{equation}

The fractions of molecules pointing into the positive and negative
hemispheres are then
\begin{align}
N_{+\unitvec{z}}
&=
\int_{0}^{2\pi}d\varphi
\int_{0}^{\pi/2}d\theta\,
\sin\theta\,|\Psi(\theta,\varphi)|^2 ,
\\
N_{-\unitvec{z}}
&=
\int_{0}^{2\pi}d\varphi
\int_{\pi/2}^{\pi}d\theta\,
\sin\theta\,|\Psi(\theta,\varphi)|^2
=
1-N_{+\unitvec{z}}.
\end{align}
Within this model, these populations are related to the orientational
first moment by
\begin{equation}
N_{+\unitvec{z}}
=
\frac{1}{2}
\left(
1+\frac{3}{2}\mathcal O_{\hat s}
\right),
\qquad
N_{-\unitvec{z}}
=
\frac{1}{2}
\left(
1-\frac{3}{2}\mathcal O_{\hat s}
\right).
\label{eq:head_tail_estimate}
\end{equation}
Thus, for example,
$\mathcal O_{\hat s}=0.19$ ($0.22$) corresponds within this approximate
angular-distribution model to
$N_{+\unitvec{z}}\simeq0.64$ ($0.67$), with the remaining
$N_{-\unitvec{z}}\simeq0.36$ ($0.33$) pointing into the opposite
hemisphere. These head--tail populations provide an intuitive,
model-dependent interpretation of the calculated orientation; the
quantity $\mathcal O_{\hat s}$ itself is the directly calculated measure
of spin--orientation locking.

\subsection{Spin-orientation correlations}
\label{subsec:correlations}

Both the orientation-conditioned measurement of mean spin direction $\langle \unitvec{s}^L\rangle_\rho$, and spin-conditioned measurement of mean molecular orientation $\langle \unitvec{e}^L \rangle_{\unitvec{s}^L}$ can be expressed as complementary first moments of the same joint distribution, i.e., 
\begin{align}
\langle \unitvec{s}^L \rangle_{\rho} = \dfrac{1}{N} \int d\Theta_p \int d\Theta_s^L \int d\Theta_k^L \, \unitvec{s}^L W^{L}(\hat{\mathbf k}^{L},\hat{\mathbf s}^{L},\rho)
\label{eq:methods-rho-conditioned}
\end{align}
\begin{align}
\langle \unitvec{e}^L \rangle_{\unitvec{s}^L}= \dfrac{1}{N} \int d\rho \int d\Theta_p \int d\Theta_k^L \, \unitvec{e}^L W^{L}(\hat{\mathbf k}^{L},\hat{\mathbf s}^{L},\rho).
\end{align}
where, $N=S_0 |E_\omega^L|^2/6$. We now define the orientation-resolved correlation tensor $G_{ij}^{L}(\rho;\unitvec{e}^M)$ as follows 
\begin{align}
G_{ij}^{L}(\rho;\unitvec{e}^M) =& \dfrac{1}{N}\int d\Theta_p \int d\Theta_s^L \int d\Theta_k^L \, e_i^{L}s_j^{L} W^{L}(\hat{\mathbf k}^{L},\hat{\mathbf s}^{L},\rho) \nonumber \\
=& e_i^L \left( \dfrac{1}{N}\int d\Theta_p \int d\Theta_s^L \int d\Theta_k^L \,s_j^{L} W^{L}(\hat{\mathbf k}^{L},\hat{\mathbf s}^{L},\rho) \right) = e_i^L \langle s_j^L \rangle_{\rho},
\end{align}
which immediately gives 
\begin{equation}
\langle \unitvec{s}^L \rangle_\rho
=
\left[
\mathbf G^{L}
\left(
\rho;\hat{\mathbf e}^{M}
\right)
\right]^T
\hat{\mathbf e}^{L}.
\end{equation}
Thus, the orientation-conditioned mean spin direction $\langle \unitvec{s}^L\rangle_\rho$ arises from the correlation tensor. 

In order to express the spin-conditioned mean molecular orientation $\langle \unitvec{e}^L \rangle_{\unitvec{s}^L}$ in terms of the correlation tensor $G_{ij}^{L}(\rho;\unitvec{e}^M)$, it will be convenient to evaluate $\langle s_i^L \rangle_{\rho}$ as follows: 
\begin{align}
\langle s_i^L \rangle_{\rho}  =& \dfrac{1}{2N} \int d\Theta_p  \int d\Theta_k^L  \left[ \sum_{I,\mu,\nu} \left( \fullvec{D}_{I,\fullvec{k}^{M},\mu}^{L*} \cdot \fullvec{E}_p^{L*}\right)\left( \fullvec{D}_{I,\fullvec{k}^{M},\nu}^{L} \cdot \fullvec{E}_p^{L}\right) \int d\Theta_s^L\, s_i^L \left( \delta_{\mu,\nu} + \unitvec{s}^L \cdot \unitvec{\sigma}_{\nu,\mu}^L \right) \right] \nonumber \\
=& \dfrac{1}{2N} \int d\Theta_p  \int d\Theta_k^L \,  \left[ \sum_{I,\mu,\nu} \left( \fullvec{D}_{I,\fullvec{k}^{M},\mu}^{L*} \cdot \fullvec{E}_p^{L*}\right)\left( \fullvec{D}_{I,\fullvec{k}^{M},\nu}^{L} \cdot \fullvec{E}_p^{L}\right) \sigma_{(\nu,\mu),j}^L \right] \int d\Theta_s \, s_i^L s_j^L \nonumber \\
=& \dfrac{1}{3} \mathcal{W}_{S,i}^L(\rho)
\end{align}
wherein, 
\begin{align}
\mathcal{W}_{S,i}^L(\rho) = \dfrac{1}{|E_\omega^L|^2 S_0} \int d\Theta_p \int d\Theta_k^L \, \text{Tr} \left[ \sum_I \left( \fullvec{D}_{I,\fullvec{k}^{M}}^{L} \cdot \fullvec{E}_p^{L}\right) \unitvec{\sigma} \left( \fullvec{D}_{I,\fullvec{k}^{M}}^{L*} \cdot \fullvec{E}_p^{L*}\right) \right]_i, 
\end{align}
which is obtained from the identities $\int d\Theta_s\,s_i=0$ and $\int d\Theta_s \, s_i s_j = \delta_{i,j}/3$. This then implies that the correlation tensor $G_{ij}^{L}(\rho;\unitvec{e}^M)$ is equivalent to 
\begin{align}
G_{ij}^{L}(\rho;\unitvec{e}^M) = \dfrac{1}{3} e_i^L \mathcal{W}_{j}^L(\rho). 
\end{align}
Explicit evaluation of $\langle \unitvec{e}^L\rangle_{\unitvec{s}^L}$ we then obtain 
\begin{align}
\langle e_i^L \rangle_{\unitvec{s}^L}=& \dfrac{1}{2N} \int d\rho \int d\Theta_p \int d\Theta_k^L \, e_i^L \left[ \sum_{I,\mu,\nu} \left( \fullvec{D}_{I,\fullvec{k}^{M},\mu}^{L*} \cdot \fullvec{E}_p^{L*}\right)\left( \fullvec{D}_{I,\fullvec{k}^{M},\nu}^{L} \cdot \fullvec{E}_p^{L}\right) \left( \delta_{\mu,\nu} + \unitvec{s}^L \cdot \unitvec{\sigma}_{\nu,\mu}^L \right) \right] \nonumber \\ 
=& \dfrac{1}{12N} \left[ \sum_{I,\mu,\nu}\delta_{\mu,\nu} e_i^M  \int d\Theta_k^M \left(  \fullvec{D}_{I,\fullvec{k}^{M},\mu}^{M*} \times \fullvec{D}_{I,\fullvec{k}^{M},\nu}^{M} \right)_i  \right] \left[ \int d\Theta_p (\fullvec{E}_p^{L*}\times\fullvec{E}_p^L)\right] \nonumber \\
&+ \left\{\dfrac{1}{2N} \int d\rho \, e_i^L \left[  \int d\Theta_p \int d\Theta_k^L \sum_{I,\mu,\nu} \left( \fullvec{D}_{I,\fullvec{k}^{M},\mu}^{L*} \cdot \fullvec{E}_p^{L*}\right)\left( \fullvec{D}_{I,\fullvec{k}^{M},\nu}^{L} \cdot \fullvec{E}_p^{L}\right) \sigma_{(\nu,\mu),j}^L \right] \right\} s_j^L \nonumber \\
=& \left\{\dfrac{3}{S_0|E_\omega^L|^2} \int d\rho \, e_i^L \left[  \int d\Theta_p \int d\Theta_k^L \sum_{I,\mu,\nu} \left( \fullvec{D}_{I,\fullvec{k}^{M},\mu}^{L*} \cdot \fullvec{E}_p^{L*}\right)\left( \fullvec{D}_{I,\fullvec{k}^{M},\nu}^{L} \cdot \fullvec{E}_p^{L}\right) \sigma_{(\nu,\mu),j}^L \right] \right\} s_j^L \nonumber \\
=& 3 \left( \int d\rho \, e_i^L \mathcal{W}_{S,j}^L(\rho)  \right) s_j^L = 3 \left( \int d\rho \, G_{ij}^{L}(\rho;\unitvec{e}^M) \right) s_j^L = 3 \langle G_{ij}^{L}(\rho;\unitvec{e}^M) \rangle \, s_j^L,
\end{align}
which also arises from the correlation tensor. 

\backmatter

\section*{Acknowledgements}

O.S., A.F.O. and P.C.F. acknowledge ERC-2021-AdG project ULISSES, grant agreement No 101054696. Views and opinions expressed are however those of the author(s) only and do not necessarily reflect those of the European Union or the European Research Council. Neither the European Union nor the granting authority can be held responsible for them. A.F.O. acknowledges funding from the Royal Society URF/R1/201333, URF/ERE/210358, and URF/ERE/231177 and from the Deutsche Forschungsgemeinschaft (DFG, German Research Foundation) - 543760364.

\begin{appendices}

\section{Chiral superpositions of Argon excited states}
\label{app:chiral-argon}

The excited states of argon have good quantum numbers $J$ and $M_J$, when considering spin–orbit interaction:
\begin{equation}
\label{eqn:true-excited-state}
\begin{aligned}
	\hat{H}_0\ket{w} &= E_w\ket{w}, &
	\hat{\vec{J}}^2\ket{w} &= J(J+1)\ket{w}, &
	\hat{J}_z\ket{w} &= M_J\ket{w}.
\end{aligned}
\end{equation}
In a close-coupling description of the excited states, the active electron is \textit{entangled} with the ion. In the particle–hole basis employed in configuration-interaction singles \cite{carlstrom2022general1,carlstrom2022general2}, the excited states have contributions from multiple ion channels:
\begin{equation}
\ket{w} =
\sum_i
\antisym
\left(\annihilate{i}\ket{\Phi_0}\right)
\ket{\chi_i},
\end{equation}
where \(\ket{\Phi_0}\) is the Hartree--Fock reference state (the ground state), \(\annihilate{i}\) annihilates the \(i\)th occupied orbital, \(\ket{\chi_i}\) is the channel-specific orbital for the excited or free electron, and \(\antisym\) is the antisymmetrization operator. \(\annihilate{i}\ket{\Phi_0}\) is thus an approximation to the state of the ion (Koopmans approximation), and the orbitals of \(\annihilate{i}\ket{\Phi_0}\) and \(\ket{\chi_i}\) are expanded in the \(\ell j m_j\) basis, wherein the spin--orbit interaction is diagonal. In our calculations, relativistic effects (of which the spin--orbit interaction is one), are treated using a relativistic effective-core potential \cite{Nicklass1995} (there are alternative approaches described in the literature \cite{Pabst2016,Zapata2022,Ruberti2024}, but for our purposes, the present approach is more suitable).

For each excited state \(\ket{w}\), we may compute the ionization dipole matrix elements \(\matrixel*{I\fullvec{k}m_s}{\hat{d}}{w}\), using a surface-flux technique \cite{Carlstroem2024spinpolspectral,carlstrom2022general1}, and they are resolved on the ion state \(I\) and the photoelectron state in terms of the magnitude \(k=\abs*{\vec{k}}\) of the linear momentum, the angular distribution in terms of the orbital angular momentum \((\ell m_\ell)_{\vec{k}}\), and the spin \(m_s\).

We however desire the ionization dipole matrix elements starting from a chiral superposition of states which as far as possible resemble factorized one-electron excited states characterized by \(n\ell m_\ell s m_s\), i.e.\ our desired initial wavefunction is given by
\begin{equation}
\ket{\Psi} =
\ket{\Psi_{\textrm{ion}}}
\left[
\sum_i
f_i\ket{(n\ell m_\ell s m_s)_i}
\right]
\defd
\ket{\Psi_{\textrm{ion}}}
\sum_i
f_i\ket{k_i},
\end{equation}
where the ionic state \(\ket{\Psi_{\textrm{ion}}}\) is shared between all terms in the expansion, and the expansion coefficients \(f_i\) may be chosen at will. In the second step, we have introduced a short-hand notation for the state of the electron.

To achieve this goal, we proceed in two steps:
\begin{enumerate}
\item Find linear combinations of the true excited states \(\ket{w}\) that are approximately factorized into an ionic part and an electronic part:
\begin{equation}
	\label{eqn:approximate-electron-states}
	\ket{\Phi_i} \approx \ket{\Phi_{\textrm{ion},i}}\ket{k_i}.
\end{equation}
There will be multiple such approximate factorizations, since there are multiple ionization channels. Below, we discuss how we try to find the \enquote{optimal} ones.
\item Given a set \(\{\ket{k_i},a_i\}\), try to find linear combinations of the factorized states \eqref{eqn:approximate-electron-states}, that simultaneously are as close to the desired chiral superposition as possible, while still maintaining maximum overlap of the ionic wavefunction, since that increases the purity of the state.
\end{enumerate}

\begin{figure}[t!]
\centering
\includegraphics[width=0.95\textwidth]{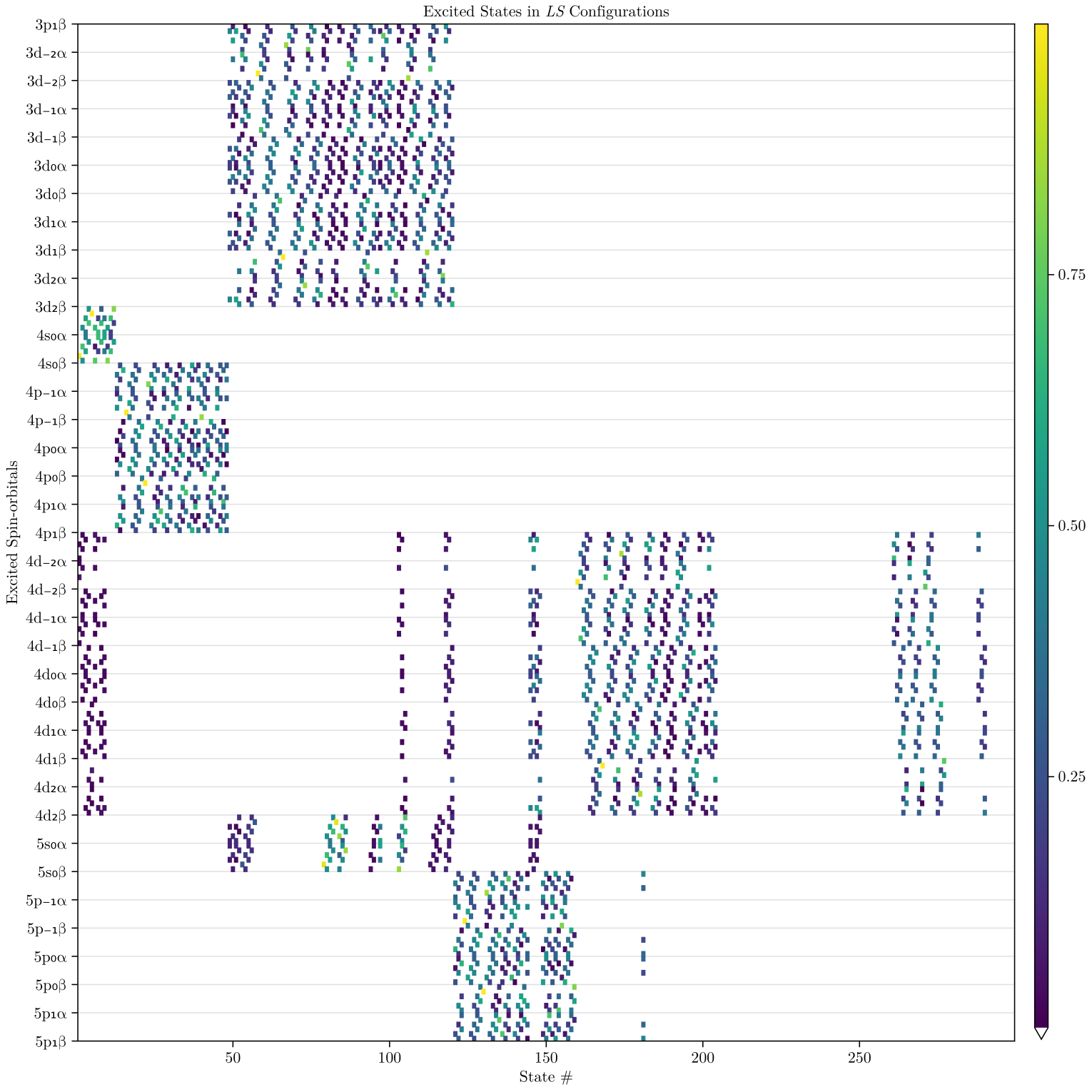}
\caption{\label{fig:configuration-overlaps}
	Overlaps between excited states and spin-configurations, i.e.\ uncoupled configurations of spin-orbitals in the \(n\ell m_\ell s m_s\) basis. The spin-configurations (the \(y\) axis) are ordered by the quantum numbers of the excited spin-orbital, given as labels pertaining to the block immediately above them. It is clearly visible that this overlap matrix is mostly composed of disjoint blocks, where the excited states are accurately classified by \(n\ell\) of the excited spin-orbital.}
\end{figure}

\subsection{State Factorization by Block Diagonalization in Orthogonal Subspaces}

We define the matrix \(\mat{B}\) with matrix elements \(b_{vw}\defd\braket{v}{w}\), which is the
projection of the true excited state \(\ket{w}\) given by
\eqref{eqn:true-excited-state} on the \textit{uncoupled}
configuration \(\ket{v}\) in the \(\ell m_\ell s m_s\) basis. We may permute the rows and columns of the
matrix \(\mat{B}\), such that it consists of mostly disjoint blocks \(\mat{B}_b\), which we can
transform separately (see Fig.~\ref{fig:configuration-overlaps}). This is possible since the quantum numbers of
the excited electron are \textit{almost} good quantum numbers. E.g.\ a \(\conf{4s_0}\alpha\) state,
will contain contributions from \(\conf{4d}_{m_\ell}\alpha\) configurations, but they will be of less
importance, compared to the dominant configurations. 

The rows in each block
\(\mat{B}_b\) are ordered such that they are grouped by the excited
orbital, and we then wish to apply a transformation to \(\mat{B}_b\)
such that it becomes approximately block diagonal, i.e.\ we find
expansions for states with a particular excited orbital (in the \(\ell
m_\ell s m_s\) basis) in terms of the true excited states (in the \(\ell j
m_j\) basis). This achieves the approximate factorization \eqref{eqn:approximate-electron-states},
thereby disentangling the states. It will however introduce an energy spread,
i.e.\ the factorized states are no longer stationary states of the
Hamiltonian.

\begin{figure}[t!]
\centering
\includegraphics[width=\textwidth]{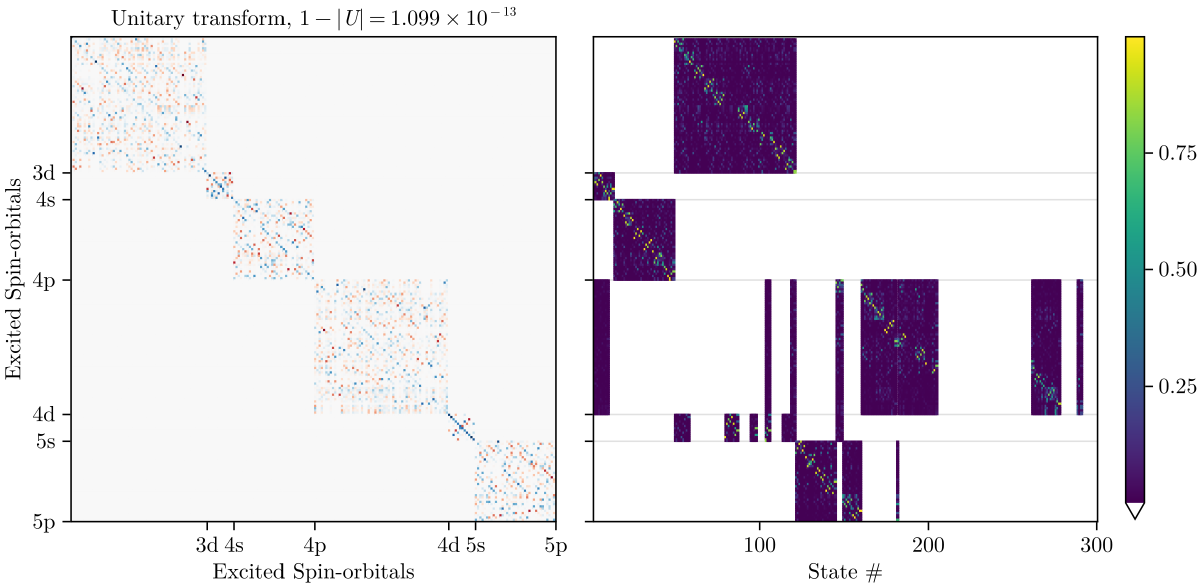}
\caption{\label{fig:unitary-transform}The left panel shows the optimized unitary transform matrix \(\mat{U}\). The right panel shows the overlap matrix \(\mat{U}\mat{B}\) between rotated excited states and the spin-configurations, again grouped by the excited spin-orbitals.}
\end{figure}

\begin{figure}[t!]
\centering
\includegraphics[width=\textwidth]{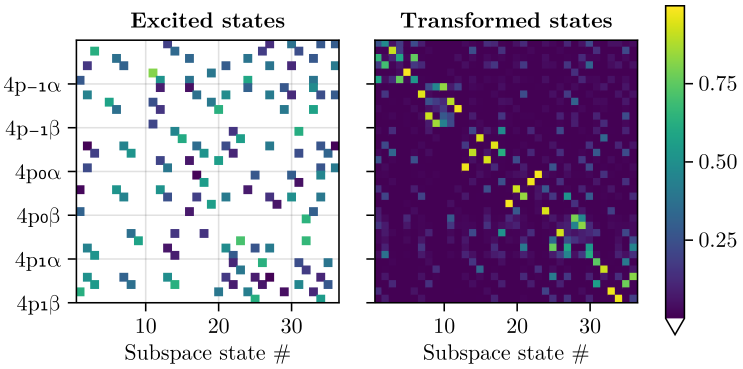}
\caption{\label{fig:transform-4p} The left panel shows the overlaps between the excited states which are dominated by excited electrons in 4p orbitals, resolved on the individual spin-orbitals. The right panel the resultant states after applying the unitary transform \(\mat{U}\) within this subspace. As can be seen, the block-diagonalization is not perfect. Additionally, we have multiple transformed states for each spin-orbital.}
\end{figure}

We consider without loss of generality a case with two different excited orbitals \(\ket{k_p}\)
and \(\ket{k_q}\) (e.g.\ \(\conf{4s_0}\spinup\) and \(\conf{4s_0}\spindown\)), and drop the
\(b\) subscript for brevity. The overlap matrix (within this block) is then given by
\begin{equation}
\mat{B} =
(\proj+\rej)
\mat{B}
(\proj+\rej),
\end{equation}
where \(\proj\) projects on the configurations containing \(\ket{k_p}\)
and \(\rej\) on configurations containing \(\ket{k_q}\). Within the chosen space, \(\proj+\rej=\identity\).
We wish to find a unitary matrix \(\mat{U}\), such that
\begin{equation}
\begin{cases}
	\proj\mat{U}\mat{B}\rej &= 0, \\
	\rej\mat{U}\mat{B}\proj &= 0,
\end{cases}
\end{equation}
i.e.\ a block-diagonalizing transform.
We will not be able to achieve this identically, therefore we
formulate it as a minimization problem:
\begin{equation}
\label{eqn:minimization-problem}
\begin{aligned}
	\min_{\mat{U}}
	&
	\norm*{\proj\mat{U}\mat{B}\rej}^2 +
	\norm*{\rej\mat{U}\mat{B}\proj}^2 \\
	\textrm{such that }
	&
	\adjoint{\mat{U}}\mat{U} = \identity.
\end{aligned}
\end{equation}
Its equivalent Lagrangian formulation, incorporating the unitarity constraint on \(\mat{U}\)
using a Lagrangian multiplier \(\lagrange\), is given by
\begin{equation}
\label{eqn:lagrangian-minimization}
\begin{aligned}
	\Lagrangian
	&=
	\tr(\adjoint{\rej}
	\adjoint{\mat{B}}
	\adjoint{\mat{U}}
	\adjoint{\proj}
	\proj
	\mat{U}
	\mat{B}
	\rej) +
	\tr(\adjoint{\proj}
	\adjoint{\mat{B}}
	\adjoint{\mat{U}}
	\adjoint{\rej}
	\rej
	\mat{U}
	\mat{B}
	\proj) +
	\lagrange{}
	[\tr(\adjoint{\mat{U}}\mat{U})-1] \\
	&=
	\tr(\adjoint{\rej}
	\adjoint{\mat{B}}
	\adjoint{\mat{U}}
	\proj
	\mat{U}
	\mat{B}
	\rej) +
	\tr(\adjoint{\proj}
	\adjoint{\mat{B}}
	\adjoint{\mat{U}}
	\rej
	\mat{U}
	\mat{B}
	\proj) +
	\lagrange{}
	[\tr(\adjoint{\mat{U}}\mat{U})-1] \\
	&=
	\tr(
	\mat{B}
	\rej
	\adjoint{\mat{B}}
	\adjoint{\mat{U}}
	\proj
	\mat{U}) +
	\tr(
	\mat{B}
	\proj
	\adjoint{\mat{B}}
	\adjoint{\mat{U}}
	\rej
	\mat{U}) +
	\lagrange{}
	[\tr(\adjoint{\mat{U}}\mat{U})-1],
\end{aligned}
\end{equation}
where we have used the cyclic property of the trace operation, the fact that \(\adjoint{\proj}\proj = \proj^2 = \proj\), and
similarly for \(\rej\). Variation of \(\Lagrangian\) with respect to \(\adjoint{\mat{U}}\),
and \(\lagrange\), respectively, yields
\begin{equation}
\label{eqn:lagrangian-variation}
\begin{aligned}
	\vary{\adjoint{\mat{U}}}\Lagrangian
	&=
	\mat{B}
	\rej
	\adjoint{\mat{B}}
	\proj
	\mat{U} +
	\mat{B}
	\proj
	\adjoint{\mat{B}}
	\rej
	\mat{U} +
	\lagrange
	\mat{U} =
	(\mat{B}
	\rej
	\adjoint{\mat{B}}
	\proj +
	\mat{B}
	\proj
	\adjoint{\mat{B}}
	\rej +
	\lagrange\identity)
	\mat{U},
	\\
	\vary{\lambda}\Lagrangian
	&=
	\adjoint{\mat{U}}\mat{U} -
	\identity,
\end{aligned}
\end{equation}
which we recognize as an eigenvalue problem. The minimization problem \eqref{eqn:lagrangian-minimization} may be solved using any
standard non-linear solver; its variation \eqref{eqn:lagrangian-variation} can be
used to improve convergence. In our implementation, however, we employ Riemannian
manifold optimization \cite{Mogensen2018optim} of the minimization problem \eqref{eqn:minimization-problem}
as-is; the matrix \(\mat{U}\) is required to stay on the \textit{Stiefel} manifold,
i.e.\ the matrix manifold of matrices with mutually orthonormal columns. In Fig.~\ref{fig:unitary-transform}, the optimized \(\mat{U}\) is shown, together with the transformed overlap matrix \(\mat{U}\mat{B}\). With this method, we are able to achieve factorizations that to \(\ge\SI{80}{\percent}\) have the excited electron in the desired spin-orbital \(n\ell m_\ell s m_s\), with some residual contamination from other states; see Fig.~\ref{fig:transform-4p} for the subspace of \(\conf{4p}\) states. The energy spread is typically small: \((\expect*{E^2}-\expect*{E}^2)^{1/2}\lesssim\num{e-2}\expect*{E}\), with the excited states of interest having energies \(\expect*{E}\gtrsim\SI{0.4}{Ha}\).

\subsection{Deriving Chiral Superpositions}

As we saw above (e.g.\ in Fig.~\ref{fig:transform-4p}), there are multiple states that have the excited electron in the same spin-orbital \(\ket{k_i}\). Any of these states may be written as a linear combination of the true excited states \eqref{eqn:true-excited-state}:
\begin{equation}
\begin{aligned}
	\ket{\Phi_i}_j &= \sum_w c^{(i)}_{jw} \ket{w} = 
	\antisym\left[\sum_w c^{(i)}_{jw} \annihilate{k_i}\ket{w}\right]\ket{k_i} &
	\implies
	\ket{\vec{\Phi}_i} &= \mat{C}_i \ket{\vec{w}} =
	\antisym\left[
	\mat{C}_i
	\annihilate{k_i}
	\ket{\vec{w}}
	\right]
	\ket{k_i},
\end{aligned}
\end{equation}
where \(\ket{\Phi_i}_j\) is the \(j\)th state with the excited electron in the spin-orbital \(\ket{k_i}\), \(c^{(i)}_{jw}\) the corresponding expansion coefficient for the true eigenstate \(\ket{w}\). We also introduce the vector notation \(\transpose{\ket{\vec{\Phi}_i}}\defd\transpose{\begin{bmatrix}\ket{\Phi_i}_1&\ket{\Phi_i}_2&\cdots\end{bmatrix}}\), and similarly for \(\ket{\vec{w}}\). \(\mat{C}_i\) are the corresponding expansion coefficients arranged into a matrix. In the second step
we have written the states on the factorized form \eqref{eqn:approximate-electron-states}.

If we now wish to create a factorized chiral superposition of the excited electron, i.e.\ we wish to define our initial state according to
\begin{equation}
\label{eqn:chiral-state-factorization}
\ket{\Psi_0} = 
\ket{\Psi_{\textrm{ion}}}
\sum_i
f_i\ket{k_i},
\end{equation}
for every \(\ket{k_i}\) we need to make a linear combination of the possible \(\ket{\Phi_i}_j\) such that the ion degrees-of-freedom maximally overlap; otherwise the factorization \eqref{eqn:chiral-state-factorization} does not hold. Symbolically, we write this as
\begin{equation}
\ket{\vec{g}_i\cdot\vec{\Phi}_i} =
\ctranspose{\vec{g}_i}
\mat{C}_i
\ket{\vec{w}} =
\antisym\left[
\ctranspose{\vec{g}_i}
\mat{C}_i
\annihilate{k_i}
\ket{\vec{w}}
\right]
\ket{k_i}.
\end{equation}
To achieve the desired factorized state \eqref{eqn:chiral-state-factorization}, we thus have to solve the following maximization problem:
\begin{equation}
\begin{aligned}
	&\max_{\mat{G}}
	\ctranspose{\left[
		\ctranspose{\vec{g}_i}
		\mat{C}_i
		\annihilate{k_i}
		\ket{\vec{w}}
		\right]}
	\left[
	\ctranspose{\vec{g}_j}
	\mat{C}_j
	\annihilate{k_j}
	\ket{\vec{w}}
	\right] =
	\max_{\mat{G}}
	\big[
	\bra{\vec{w}}
	\create{k_i}
	\ctranspose{\mat{C}_i}
	\vec{g}_i
	\big]
	\big[
	\ctranspose{\vec{g}_j}
	\mat{C}_j
	\annihilate{k_j}
	\ket{\vec{w}}
	\big],\\
	&\textrm{such that }
	\norm{\vec{g}_i}^2 = 1, \forall i,
\end{aligned}
\end{equation}
where the \(i\)th column of the matrix \(\mat{G}\) is the vector \(\vec{g}_i\).

\end{appendices}

\bibliography{spin.bib}% common bib file

\end{document}